
\catcode`\@=11


\message{Loading jyTeX fonts...}



\font\vptrm=cmr5 \font\vptmit=cmmi5 \font\vptsy=cmsy5 \font\vptbf=cmbx5

\skewchar\vptmit='177 \skewchar\vptsy='60 \fontdimen16
\vptsy=\the\fontdimen17 \vptsy

\def\vpt{\ifmmode\err@badsizechange\else
     \@mathfontinit
     \textfont0=\vptrm  \scriptfont0=\vptrm  \scriptscriptfont0=\vptrm
     \textfont1=\vptmit \scriptfont1=\vptmit \scriptscriptfont1=\vptmit
     \textfont2=\vptsy  \scriptfont2=\vptsy  \scriptscriptfont2=\vptsy
     \textfont3=\xptex  \scriptfont3=\xptex  \scriptscriptfont3=\xptex
     \textfont\bffam=\vptbf
     \scriptfont\bffam=\vptbf
     \scriptscriptfont\bffam=\vptbf
     \@fontstyleinit
     \def\rm{\vptrm\fam=\z@}%
     \def\bf{\vptbf\fam=\bffam}%
     \def\oldstyle{\vptmit\fam=\@ne}%
     \rm\fi}


\font\viptrm=cmr6 \font\viptmit=cmmi6 \font\viptsy=cmsy6
\font\viptbf=cmbx6

\skewchar\viptmit='177 \skewchar\viptsy='60 \fontdimen16
\viptsy=\the\fontdimen17 \viptsy

\def\vipt{\ifmmode\err@badsizechange\else
     \@mathfontinit
     \textfont0=\viptrm  \scriptfont0=\vptrm  \scriptscriptfont0=\vptrm
     \textfont1=\viptmit \scriptfont1=\vptmit \scriptscriptfont1=\vptmit
     \textfont2=\viptsy  \scriptfont2=\vptsy  \scriptscriptfont2=\vptsy
     \textfont3=\xptex   \scriptfont3=\xptex  \scriptscriptfont3=\xptex
     \textfont\bffam=\viptbf
     \scriptfont\bffam=\vptbf
     \scriptscriptfont\bffam=\vptbf
     \@fontstyleinit
     \def\rm{\viptrm\fam=\z@}%
     \def\bf{\viptbf\fam=\bffam}%
     \def\oldstyle{\viptmit\fam=\@ne}%
     \rm\fi}

\font\viiptrm=cmr7 \font\viiptmit=cmmi7 \font\viiptsy=cmsy7
\font\viiptit=cmti7 \font\viiptbf=cmbx7

\skewchar\viiptmit='177 \skewchar\viiptsy='60 \fontdimen16
\viiptsy=\the\fontdimen17 \viiptsy

\def\viipt{\ifmmode\err@badsizechange\else
     \@mathfontinit
     \textfont0=\viiptrm  \scriptfont0=\vptrm  \scriptscriptfont0=\vptrm
     \textfont1=\viiptmit \scriptfont1=\vptmit \scriptscriptfont1=\vptmit
     \textfont2=\viiptsy  \scriptfont2=\vptsy  \scriptscriptfont2=\vptsy
     \textfont3=\xptex    \scriptfont3=\xptex  \scriptscriptfont3=\xptex
     \textfont\itfam=\viiptit
     \scriptfont\itfam=\viiptit
     \scriptscriptfont\itfam=\viiptit
     \textfont\bffam=\viiptbf
     \scriptfont\bffam=\vptbf
     \scriptscriptfont\bffam=\vptbf
     \@fontstyleinit
     \def\rm{\viiptrm\fam=\z@}%
     \def\it{\viiptit\fam=\itfam}%
     \def\bf{\viiptbf\fam=\bffam}%
     \def\oldstyle{\viiptmit\fam=\@ne}%
     \rm\fi}


\font\viiiptrm=cmr8 \font\viiiptmit=cmmi8 \font\viiiptsy=cmsy8
\font\viiiptit=cmti8
\font\viiiptbf=cmbx8

\skewchar\viiiptmit='177 \skewchar\viiiptsy='60 \fontdimen16
\viiiptsy=\the\fontdimen17 \viiiptsy

\def\viiipt{\ifmmode\err@badsizechange\else
     \@mathfontinit
     \textfont0=\viiiptrm  \scriptfont0=\viptrm  \scriptscriptfont0=\vptrm
     \textfont1=\viiiptmit \scriptfont1=\viptmit \scriptscriptfont1=\vptmit
     \textfont2=\viiiptsy  \scriptfont2=\viptsy  \scriptscriptfont2=\vptsy
     \textfont3=\xptex     \scriptfont3=\xptex   \scriptscriptfont3=\xptex
     \textfont\itfam=\viiiptit
     \scriptfont\itfam=\viiptit
     \scriptscriptfont\itfam=\viiptit
     \textfont\bffam=\viiiptbf
     \scriptfont\bffam=\viptbf
     \scriptscriptfont\bffam=\vptbf
     \@fontstyleinit
     \def\rm{\viiiptrm\fam=\z@}%
     \def\it{\viiiptit\fam=\itfam}%
     \def\bf{\viiiptbf\fam=\bffam}%
     \def\oldstyle{\viiiptmit\fam=\@ne}%
     \rm\fi}


\def\getixpt{%
     \font\ixptrm=cmr9
     \font\ixptmit=cmmi9
     \font\ixptsy=cmsy9
     \font\ixptit=cmti9
     \font\ixptbf=cmbx9
     \skewchar\ixptmit='177 \skewchar\ixptsy='60
     \fontdimen16 \ixptsy=\the\fontdimen17 \ixptsy}

\def\ixpt{\ifmmode\err@badsizechange\else
     \@mathfontinit
     \textfont0=\ixptrm  \scriptfont0=\viiptrm  \scriptscriptfont0=\vptrm
     \textfont1=\ixptmit \scriptfont1=\viiptmit \scriptscriptfont1=\vptmit
     \textfont2=\ixptsy  \scriptfont2=\viiptsy  \scriptscriptfont2=\vptsy
     \textfont3=\xptex   \scriptfont3=\xptex    \scriptscriptfont3=\xptex
     \textfont\itfam=\ixptit
     \scriptfont\itfam=\viiptit
     \scriptscriptfont\itfam=\viiptit
     \textfont\bffam=\ixptbf
     \scriptfont\bffam=\viiptbf
     \scriptscriptfont\bffam=\vptbf
     \@fontstyleinit
     \def\rm{\ixptrm\fam=\z@}%
     \def\it{\ixptit\fam=\itfam}%
     \def\bf{\ixptbf\fam=\bffam}%
     \def\oldstyle{\ixptmit\fam=\@ne}%
     \rm\fi}


\font\xptrm=cmr10 \font\xptmit=cmmi10 \font\xptsy=cmsy10
\font\xptex=cmex10 \font\xptit=cmti10 \font\xptsl=cmsl10
\font\xptbf=cmbx10 \font\xpttt=cmtt10 \font\xptss=cmss10
\font\xptsc=cmcsc10 \font\xptbfs=cmb10 \font\xptbmit=cmmib10

\skewchar\xptmit='177 \skewchar\xptbmit='177 \skewchar\xptsy='60
\fontdimen16 \xptsy=\the\fontdimen17 \xptsy

\def\xpt{\ifmmode\err@badsizechange\else
     \@mathfontinit
     \textfont0=\xptrm  \scriptfont0=\viiptrm  \scriptscriptfont0=\vptrm
     \textfont1=\xptmit \scriptfont1=\viiptmit \scriptscriptfont1=\vptmit
     \textfont2=\xptsy  \scriptfont2=\viiptsy  \scriptscriptfont2=\vptsy
     \textfont3=\xptex  \scriptfont3=\xptex    \scriptscriptfont3=\xptex
     \textfont\itfam=\xptit
     \scriptfont\itfam=\viiptit
     \scriptscriptfont\itfam=\viiptit
     \textfont\bffam=\xptbf
     \scriptfont\bffam=\viiptbf
     \scriptscriptfont\bffam=\vptbf
     \textfont\bfsfam=\xptbfs
     \scriptfont\bfsfam=\viiptbf
     \scriptscriptfont\bfsfam=\vptbf
     \textfont\bmitfam=\xptbmit
     \scriptfont\bmitfam=\viiptmit
     \scriptscriptfont\bmitfam=\vptmit
     \@fontstyleinit
     \def\rm{\xptrm\fam=\z@}%
     \def\it{\xptit\fam=\itfam}%
     \def\sl{\xptsl}%
     \def\bf{\xptbf\fam=\bffam}%
     \def\tt{\xpttt}%
     \def\ss{\xptss}%
     \def\sc{\xptsc}%
     \def\bfs{\xptbfs\fam=\bfsfam}%
     \def\bmit{\fam=\bmitfam}%
     \def\oldstyle{\xptmit\fam=\@ne}%
     \rm\fi}


\def\getxipt{%
     \font\xiptrm=cmr10  scaled\magstephalf
     \font\xiptmit=cmmi10 scaled\magstephalf
     \font\xiptsy=cmsy10 scaled\magstephalf
     \font\xiptex=cmex10 scaled\magstephalf
     \font\xiptit=cmti10 scaled\magstephalf
     \font\xiptsl=cmsl10 scaled\magstephalf
     \font\xiptbf=cmbx10 scaled\magstephalf
     \font\xipttt=cmtt10 scaled\magstephalf
     \font\xiptss=cmss10 scaled\magstephalf
     \skewchar\xiptmit='177 \skewchar\xiptsy='60
     \fontdimen16 \xiptsy=\the\fontdimen17 \xiptsy}

\def\xipt{\ifmmode\err@badsizechange\else
     \@mathfontinit
     \textfont0=\xiptrm  \scriptfont0=\viiiptrm  \scriptscriptfont0=\viptrm
     \textfont1=\xiptmit \scriptfont1=\viiiptmit \scriptscriptfont1=\viptmit
     \textfont2=\xiptsy  \scriptfont2=\viiiptsy  \scriptscriptfont2=\viptsy
     \textfont3=\xiptex  \scriptfont3=\xptex     \scriptscriptfont3=\xptex
     \textfont\itfam=\xiptit
     \scriptfont\itfam=\viiiptit
     \scriptscriptfont\itfam=\viiptit
     \textfont\bffam=\xiptbf
     \scriptfont\bffam=\viiiptbf
     \scriptscriptfont\bffam=\viptbf
     \@fontstyleinit
     \def\rm{\xiptrm\fam=\z@}%
     \def\it{\xiptit\fam=\itfam}%
     \def\sl{\xiptsl}%
     \def\bf{\xiptbf\fam=\bffam}%
     \def\tt{\xipttt}%
     \def\ss{\xiptss}%
     \def\oldstyle{\xiptmit\fam=\@ne}%
     \rm\fi}


\font\xiiptrm=cmr12 \font\xiiptmit=cmmi12 \font\xiiptsy=cmsy10
scaled\magstep1 \font\xiiptex=cmex10  scaled\magstep1
\font\xiiptit=cmti12 \font\xiiptsl=cmsl12 \font\xiiptbf=cmbx12
\font\xiiptss=cmss12 \font\xiiptsc=cmcsc10 scaled\magstep1
\font\xiiptbfs=cmb10  scaled\magstep1 \font\xiiptbmit=cmmib10
scaled\magstep1

\skewchar\xiiptmit='177 \skewchar\xiiptbmit='177 \skewchar\xiiptsy='60
\fontdimen16 \xiiptsy=\the\fontdimen17 \xiiptsy

\def\xiipt{\ifmmode\err@badsizechange\else
     \@mathfontinit
     \textfont0=\xiiptrm  \scriptfont0=\viiiptrm  \scriptscriptfont0=\viptrm
     \textfont1=\xiiptmit \scriptfont1=\viiiptmit \scriptscriptfont1=\viptmit
     \textfont2=\xiiptsy  \scriptfont2=\viiiptsy  \scriptscriptfont2=\viptsy
     \textfont3=\xiiptex  \scriptfont3=\xptex     \scriptscriptfont3=\xptex
     \textfont\itfam=\xiiptit
     \scriptfont\itfam=\viiiptit
     \scriptscriptfont\itfam=\viiptit
     \textfont\bffam=\xiiptbf
     \scriptfont\bffam=\viiiptbf
     \scriptscriptfont\bffam=\viptbf
     \textfont\bfsfam=\xiiptbfs
     \scriptfont\bfsfam=\viiiptbf
     \scriptscriptfont\bfsfam=\viptbf
     \textfont\bmitfam=\xiiptbmit
     \scriptfont\bmitfam=\viiiptmit
     \scriptscriptfont\bmitfam=\viptmit
     \@fontstyleinit
     \def\rm{\xiiptrm\fam=\z@}%
     \def\it{\xiiptit\fam=\itfam}%
     \def\sl{\xiiptsl}%
     \def\bf{\xiiptbf\fam=\bffam}%
     \def\tt{\xiipttt}%
     \def\ss{\xiiptss}%
     \def\sc{\xiiptsc}%
     \def\bfs{\xiiptbfs\fam=\bfsfam}%
     \def\bmit{\fam=\bmitfam}%
     \def\oldstyle{\xiiptmit\fam=\@ne}%
     \rm\fi}


\def\getxiiipt{%
     \font\xiiiptrm=cmr12  scaled\magstephalf
     \font\xiiiptmit=cmmi12 scaled\magstephalf
     \font\xiiiptsy=cmsy9  scaled\magstep2
     \font\xiiiptit=cmti12 scaled\magstephalf
     \font\xiiiptsl=cmsl12 scaled\magstephalf
     \font\xiiiptbf=cmbx12 scaled\magstephalf
     \font\xiiipttt=cmtt12 scaled\magstephalf
     \font\xiiiptss=cmss12 scaled\magstephalf
     \skewchar\xiiiptmit='177 \skewchar\xiiiptsy='60
     \fontdimen16 \xiiiptsy=\the\fontdimen17 \xiiiptsy}

\def\xiiipt{\ifmmode\err@badsizechange\else
     \@mathfontinit
     \textfont0=\xiiiptrm  \scriptfont0=\xptrm  \scriptscriptfont0=\viiptrm
     \textfont1=\xiiiptmit \scriptfont1=\xptmit \scriptscriptfont1=\viiptmit
     \textfont2=\xiiiptsy  \scriptfont2=\xptsy  \scriptscriptfont2=\viiptsy
     \textfont3=\xivptex   \scriptfont3=\xptex  \scriptscriptfont3=\xptex
     \textfont\itfam=\xiiiptit
     \scriptfont\itfam=\xptit
     \scriptscriptfont\itfam=\viiptit
     \textfont\bffam=\xiiiptbf
     \scriptfont\bffam=\xptbf
     \scriptscriptfont\bffam=\viiptbf
     \@fontstyleinit
     \def\rm{\xiiiptrm\fam=\z@}%
     \def\it{\xiiiptit\fam=\itfam}%
     \def\sl{\xiiiptsl}%
     \def\bf{\xiiiptbf\fam=\bffam}%
     \def\tt{\xiiipttt}%
     \def\ss{\xiiiptss}%
     \def\oldstyle{\xiiiptmit\fam=\@ne}%
     \rm\fi}


\font\xivptrm=cmr12   scaled\magstep1 \font\xivptmit=cmmi12
scaled\magstep1 \font\xivptsy=cmsy10  scaled\magstep2
\font\xivptex=cmex10  scaled\magstep2 \font\xivptit=cmti12
scaled\magstep1 \font\xivptsl=cmsl12  scaled\magstep1
\font\xivptbf=cmbx12  scaled\magstep1
\font\xivptss=cmss12  scaled\magstep1 \font\xivptsc=cmcsc10
scaled\magstep2 \font\xivptbfs=cmb10  scaled\magstep2
\font\xivptbmit=cmmib10 scaled\magstep2

\skewchar\xivptmit='177 \skewchar\xivptbmit='177 \skewchar\xivptsy='60
\fontdimen16 \xivptsy=\the\fontdimen17 \xivptsy

\def\xivpt{\ifmmode\err@badsizechange\else
     \@mathfontinit
     \textfont0=\xivptrm  \scriptfont0=\xptrm  \scriptscriptfont0=\viiptrm
     \textfont1=\xivptmit \scriptfont1=\xptmit \scriptscriptfont1=\viiptmit
     \textfont2=\xivptsy  \scriptfont2=\xptsy  \scriptscriptfont2=\viiptsy
     \textfont3=\xivptex  \scriptfont3=\xptex  \scriptscriptfont3=\xptex
     \textfont\itfam=\xivptit
     \scriptfont\itfam=\xptit
     \scriptscriptfont\itfam=\viiptit
     \textfont\bffam=\xivptbf
     \scriptfont\bffam=\xptbf
     \scriptscriptfont\bffam=\viiptbf
     \textfont\bfsfam=\xivptbfs
     \scriptfont\bfsfam=\xptbfs
     \scriptscriptfont\bfsfam=\viiptbf
     \textfont\bmitfam=\xivptbmit
     \scriptfont\bmitfam=\xptbmit
     \scriptscriptfont\bmitfam=\viiptmit
     \@fontstyleinit
     \def\rm{\xivptrm\fam=\z@}%
     \def\it{\xivptit\fam=\itfam}%
     \def\sl{\xivptsl}%
     \def\bf{\xivptbf\fam=\bffam}%
     \def\tt{\xivpttt}%
     \def\ss{\xivptss}%
     \def\sc{\xivptsc}%
     \def\bfs{\xivptbfs\fam=\bfsfam}%
     \def\bmit{\fam=\bmitfam}%
     \def\oldstyle{\xivptmit\fam=\@ne}%
     \rm\fi}


\font\xviiptrm=cmr17 \font\xviiptmit=cmmi12 scaled\magstep2
\font\xviiptsy=cmsy10 scaled\magstep3 \font\xviiptex=cmex10
scaled\magstep3 \font\xviiptit=cmti12 scaled\magstep2
\font\xviiptbf=cmbx12 scaled\magstep2 \font\xviiptbfs=cmb10
scaled\magstep3

\skewchar\xviiptmit='177 \skewchar\xviiptsy='60 \fontdimen16
\xviiptsy=\the\fontdimen17 \xviiptsy

\def\xviipt{\ifmmode\err@badsizechange\else
     \@mathfontinit
     \textfont0=\xviiptrm  \scriptfont0=\xiiptrm  \scriptscriptfont0=\viiiptrm
     \textfont1=\xviiptmit \scriptfont1=\xiiptmit \scriptscriptfont1=\viiiptmit
     \textfont2=\xviiptsy  \scriptfont2=\xiiptsy  \scriptscriptfont2=\viiiptsy
     \textfont3=\xviiptex  \scriptfont3=\xiiptex  \scriptscriptfont3=\xptex
     \textfont\itfam=\xviiptit
     \scriptfont\itfam=\xiiptit
     \scriptscriptfont\itfam=\viiiptit
     \textfont\bffam=\xviiptbf
     \scriptfont\bffam=\xiiptbf
     \scriptscriptfont\bffam=\viiiptbf
     \textfont\bfsfam=\xviiptbfs
     \scriptfont\bfsfam=\xiiptbfs
     \scriptscriptfont\bfsfam=\viiiptbf
     \@fontstyleinit
     \def\rm{\xviiptrm\fam=\z@}%
     \def\it{\xviiptit\fam=\itfam}%
     \def\bf{\xviiptbf\fam=\bffam}%
     \def\bfs{\xviiptbfs\fam=\bfsfam}%
     \def\oldstyle{\xviiptmit\fam=\@ne}%
     \rm\fi}


\font\xxiptrm=cmr17  scaled\magstep1


\def\xxipt{\ifmmode\err@badsizechange\else
     \@mathfontinit
     \@fontstyleinit
     \def\rm{\xxiptrm\fam=\z@}%
     \rm\fi}


\font\xxvptrm=cmr17  scaled\magstep2


\def\xxvpt{\ifmmode\err@badsizechange\else
     \@mathfontinit
     \@fontstyleinit
     \def\rm{\xxvptrm\fam=\z@}%
     \rm\fi}




\message{Loading jyTeX macros...}

\message{modifications to plain.tex,}


\def\newcount{\alloc@0\count\countdef\insc@unt}
\def\newdimen{\alloc@1\dimen\dimendef\insc@unt}
\def\newskip{\alloc@2\skip\skipdef\insc@unt}
\def\newmuskip{\alloc@3\muskip\muskipdef\@cclvi}
\def\newbox{\alloc@4\box\chardef\insc@unt}
\def\newtoks{\alloc@5\toks\toksdef\@cclvi}
\def\newhelp#1#2{\newtoks#1\global#1\expandafter{\csname#2\endcsname}}
\def\newread{\alloc@6\read\chardef\sixt@@n}
\def\newwrite{\alloc@7\write\chardef\sixt@@n}
\def\newfam{\alloc@8\fam\chardef\sixt@@n}
\def\newinsert#1{\global\advance\insc@unt by\m@ne
     \ch@ck0\insc@unt\count
     \ch@ck1\insc@unt\dimen
     \ch@ck2\insc@unt\skip
     \ch@ck4\insc@unt\box
     \allocationnumber=\insc@unt
     \global\chardef#1=\allocationnumber
     \wlog{\string#1=\string\insert\the\allocationnumber}}
\def\newif#1{\count@\escapechar \escapechar\m@ne
     \expandafter\expandafter\expandafter
          \xdef\@if#1{true}{\let\noexpand#1=\noexpand\iftrue}%
     \expandafter\expandafter\expandafter
          \xdef\@if#1{false}{\let\noexpand#1=\noexpand\iffalse}%
     \global\@if#1{false}\escapechar=\count@}


\newlinechar=`\^^J
\overfullrule=0pt




\let\itfam=\undefined

\let\bffam=\undefined

\count18=3


\chardef\sharps="19


\mathchardef\alpha="710B \mathchardef\beta="710C \mathchardef\gamma="710D
\mathchardef\delta="710E \mathchardef\epsilon="710F
\mathchardef\zeta="7110 \mathchardef\eta="7111 \mathchardef\theta="7112
\mathchardef\iota="7113 \mathchardef\kappa="7114
\mathchardef\lambda="7115 \mathchardef\mu="7116 \mathchardef\nu="7117
\mathchardef\xi="7118 \mathchardef\pi="7119 \mathchardef\rho="711A
\mathchardef\sigma="711B \mathchardef\tau="711C
\mathchardef\upsilon="711D \mathchardef\phi="711E \mathchardef\chi="711F
\mathchardef\psi="7120 \mathchardef\omega="7121
\mathchardef\varepsilon="7122 \mathchardef\vartheta="7123
\mathchardef\varpi="7124 \mathchardef\varrho="7125
\mathchardef\varsigma="7126 \mathchardef\varphi="7127
\mathchardef\imath="717B \mathchardef\jmath="717C \mathchardef\ell="7160
\mathchardef\wp="717D \mathchardef\partial="7140 \mathchardef\flat="715B
\mathchardef\natural="715C \mathchardef\sharp="715D



\def\angle{{\vbox{\ialign{$\m@th\scriptstyle##$\crcr
     \not\mathrel{\mkern14mu}\crcr
     \noalign{\nointerlineskip}
     \mkern2.5mu\leaders\hrule height.34\rp@\hfill\mkern2.5mu\crcr}}}}
\def\vdots{\vbox{\baselineskip4\rp@ \lineskiplimit\z@
     \kern6\rp@\hbox{.}\hbox{.}\hbox{.}}}
\def\ddots{\mathinner{\mkern1mu\raise7\rp@\vbox{\kern7\rp@\hbox{.}}\mkern2mu
     \raise4\rp@\hbox{.}\mkern2mu\raise\rp@\hbox{.}\mkern1mu}}
\def\overrightarrow#1{\vbox{\ialign{##\crcr
     \rightarrowfill\crcr
     \noalign{\kern-\rp@\nointerlineskip}
     $\hfil\displaystyle{#1}\hfil$\crcr}}}
\def\overleftarrow#1{\vbox{\ialign{##\crcr
     \leftarrowfill\crcr
     \noalign{\kern-\rp@\nointerlineskip}
     $\hfil\displaystyle{#1}\hfil$\crcr}}}
\def\overbrace#1{\mathop{\vbox{\ialign{##\crcr
     \noalign{\kern3\rp@}
     \downbracefill\crcr
     \noalign{\kern3\rp@\nointerlineskip}
     $\hfil\displaystyle{#1}\hfil$\crcr}}}\limits}
\def\underbrace#1{\mathop{\vtop{\ialign{##\crcr
     $\hfil\displaystyle{#1}\hfil$\crcr
     \noalign{\kern3\rp@\nointerlineskip}
     \upbracefill\crcr
     \noalign{\kern3\rp@}}}}\limits}
\def\big#1{{\hbox{$\left#1\vbox to8.5\rp@ {}\right.\n@space$}}}
\def\Big#1{{\hbox{$\left#1\vbox to11.5\rp@ {}\right.\n@space$}}}
\def\bigg#1{{\hbox{$\left#1\vbox to14.5\rp@ {}\right.\n@space$}}}
\def\Bigg#1{{\hbox{$\left#1\vbox to17.5\rp@ {}\right.\n@space$}}}
\def\@vereq#1#2{\lower.5\rp@\vbox{\baselineskip\z@skip\lineskip-.5\rp@
     \ialign{$\m@th#1\hfil##\hfil$\crcr#2\crcr=\crcr}}}
\def\rlh@#1{\vcenter{\hbox{\ooalign{\raise2\rp@
     \hbox{$#1\rightharpoonup$}\crcr
     $#1\leftharpoondown$}}}}
\def\bordermatrix#1{\begingroup\m@th
     \setbox\z@\vbox{%
          \def\cr{\crcr\noalign{\kern2\rp@\global\let\cr\endline}}%
          \ialign{$##$\hfil\kern2\rp@\kern\p@renwd
               &\thinspace\hfil$##$\hfil&&\quad\hfil$##$\hfil\crcr
               \omit\strut\hfil\crcr
               \noalign{\kern-\baselineskip}%
               #1\crcr\omit\strut\cr}}%
     \setbox\tw@\vbox{\unvcopy\z@\global\setbox\@ne\lastbox}%
     \setbox\tw@\hbox{\unhbox\@ne\unskip\global\setbox\@ne\lastbox}%
     \setbox\tw@\hbox{$\kern\wd\@ne\kern-\p@renwd\left(\kern-\wd\@ne
          \global\setbox\@ne\vbox{\box\@ne\kern2\rp@}%
          \vcenter{\kern-\ht\@ne\unvbox\z@\kern-\baselineskip}%
          \,\right)$}%
     \null\;\vbox{\kern\ht\@ne\box\tw@}\endgroup}
\def\endinsert{\egroup
     \if@mid\dimen@\ht\z@
          \advance\dimen@\dp\z@
          \advance\dimen@12\rp@
          \advance\dimen@\pagetotal
          \ifdim\dimen@>\pagegoal\@midfalse\p@gefalse\fi
     \fi
     \if@mid\bigskip\box\z@
          \bigbreak
     \else\insert\topins{\penalty100 \splittopskip\z@skip
               \splitmaxdepth\maxdimen\floatingpenalty\z@
               \ifp@ge\dimen@\dp\z@
                    \vbox to\vsize{\unvbox\z@\kern-\dimen@}%
               \else\box\z@\nobreak\bigskip
               \fi}%
     \fi
     \endgroup}


\def\cases#1{\left\{\,\vcenter{\m@th
     \ialign{$##\hfil$&\quad##\hfil\crcr#1\crcr}}\right.}
\def\matrix#1{\null\,\vcenter{\m@th
     \ialign{\hfil$##$\hfil&&\quad\hfil$##$\hfil\crcr
          \mathstrut\crcr
          \noalign{\kern-\baselineskip}
          #1\crcr
          \mathstrut\crcr
          \noalign{\kern-\baselineskip}}}\,}


\newif\ifraggedbottom

\def\raggedbottom{\ifraggedbottom\else
     \advance\topskip by\z@ plus60pt \raggedbottomtrue\fi}%
\def\normalbottom{\ifraggedbottom
     \advance\topskip by\z@ plus-60pt \raggedbottomfalse\fi}

\message{hacks,}


\toksdef\toks@i=1 \toksdef\toks@ii=2


\def\TeX{T\kern-.1667em \lower.5ex \hbox{E}\kern-.125em X\null}
\def\jyTeX{{\leavevmode
     \raise.587ex \hbox{\it\j}\kern-.1em \lower.048ex \hbox{\it y}\kern-.12em
     \TeX}}

\let\then=\iftrue
\def\ifnoarg#1\then{\def\hack@{#1}\ifx\hack@\empty}
\def\ifundefined#1\then{%
     \expandafter\ifx\csname\expandafter\blank\string#1\endcsname\relax}
\def\useif#1\then{\csname#1\endcsname}
\def\usename#1{\csname#1\endcsname}
\def\useafter#1#2{\expandafter#1\csname#2\endcsname}

\long\def\loop#1\repeat{\def\@iterate{#1\expandafter\@iterate\fi}\@iterate
     \let\@iterate=\relax}

\let\TeXend=\end
\def\begin#1{\begingroup\def\@@blockname{#1}\usename{begin#1}}
\def\end#1{\usename{end#1}\def\hack@{#1}%
     \ifx\@@blockname\hack@
          \endgroup
     \else\err@badgroup\hack@\@@blockname
     \fi}
\def\@@blockname{}

\def\defaultoption[#1]#2{%
     \def\hack@{\ifx\hack@ii[\toks@={#2}\else\toks@={#2[#1]}\fi\the\toks@}%
     \futurelet\hack@ii\hack@}

\def\markup#1{\let\@@marksf=\empty
     \ifhmode\edef\@@marksf{\spacefactor=\the\spacefactor\relax}\/\fi
     ${}^{\hbox{\subscriptfonts#1}}$\@@marksf}


\newtoks\shortyear
\newtoks\militaryhour
\newtoks\standardhour
\newtoks\minute
\newtoks\amorpm

\def\settime{\count@=\time\divide\count@ by60
     \militaryhour=\expandafter{\number\count@}%
     {\multiply\count@ by-60 \advance\count@ by\time
          \xdef\hack@{\ifnum\count@<10 0\fi\number\count@}}%
     \minute=\expandafter{\hack@}%
     \ifnum\count@<12
          \amorpm={am}
     \else\amorpm={pm}
          \ifnum\count@>12 \advance\count@ by-12 \fi
     \fi
     \standardhour=\expandafter{\number\count@}%
     \def\hack@19##1##2{\shortyear={##1##2}}%
          \expandafter\hack@\the\year}

\def\monthword#1{%
     \ifcase#1
          $\bullet$\err@badcountervalue{monthword}%
          \or January\or February\or March\or April\or May\or June%
          \or July\or August\or September\or October\or November\or December%
     \else$\bullet$\err@badcountervalue{monthword}%
     \fi}

\def\monthabbr#1{%
     \ifcase#1
          $\bullet$\err@badcountervalue{monthabbr}%
          \or Jan\or Feb\or Mar\or Apr\or May\or Jun%
          \or Jul\or Aug\or Sep\or Oct\or Nov\or Dec%
     \else$\bullet$\err@badcountervalue{monthabbr}%
     \fi}

\def\militarytime{\the\militaryhour:\the\minute}
\def\standardtime{\the\standardhour:\the\minute}


\def\@setnumstyle#1#2{\expandafter\global\expandafter\expandafter
     \expandafter\let\expandafter\expandafter
     \csname @\expandafter\blank\string#1style\endcsname
     \csname#2\endcsname}
\def\numstyle#1{\usename{@\expandafter\blank\string#1style}#1}
\def\ifblank#1\then{\useafter\ifx{@\expandafter\blank\string#1}\blank}

\def\blank#1{}

\def\Roman#1{\expandafter\uppercase\expandafter{\romannumeral#1}}
\def\alphabetic#1{%
     \ifcase#1
          $\bullet$\err@badcountervalue{alphabetic}%
          \or a\or b\or c\or d\or e\or f\or g\or h\or i\or j\or k\or l\or m%
          \or n\or o\or p\or q\or r\or s\or t\or u\or v\or w\or x\or y\or z%
     \else$\bullet$\err@badcountervalue{alphabetic}%
     \fi}
\def\Alphabetic#1{\expandafter\uppercase\expandafter{\alphabetic{#1}}}
\def\symbols#1{%
     \ifcase#1
          $\bullet$\err@badcountervalue{symbols}%
          \or*\or\dag\or\ddag\or\S\or$\|$%
          \or**\or\dag\dag\or\ddag\ddag\or\S\S\or$\|\|$%
     \else$\bullet$\err@badcountervalue{symbols}%
     \fi}


\catcode`\^^?=13 \def^^?{\relax}

\def\trimleading#1\to#2{\edef#2{#1}%
     \expandafter\@trimleading\expandafter#2#2^^?^^?}
\def\@trimleading#1#2#3^^?{\ifx#2^^?\def#1{}\else\def#1{#2#3}\fi}

\def\trimtrailing#1\to#2{\edef#2{#1}%
     \expandafter\@trimtrailing\expandafter#2#2^^? ^^?\relax}
\def\@trimtrailing#1#2 ^^?#3{\ifx#3\relax\toks@={}%
     \else\def#1{#2}\toks@={\trimtrailing#1\to#1}\fi
     \the\toks@}

\def\trim#1\to#2{\trimleading#1\to#2\trimtrailing#2\to#2}

\catcode`\^^?=15


\long\def\additemL#1\to#2{\toks@={\^^\{#1}}\toks@ii=\expandafter{#2}%
     \xdef#2{\the\toks@\the\toks@ii}}

\long\def\additemR#1\to#2{\toks@={\^^\{#1}}\toks@ii=\expandafter{#2}%
     \xdef#2{\the\toks@ii\the\toks@}}

\def\getitemL#1\to#2{\expandafter\@getitemL#1\hack@#1#2}
\def\@getitemL\^^\#1#2\hack@#3#4{\def#4{#1}\def#3{#2}}

\message{font macros,}


\newdimen\rp@
\newcount\@@sizeindex \@@sizeindex=0
\newcount\@@factori
\newcount\@@factorii
\newcount\@@factoriii
\newcount\@@factoriv

\countdef\maxfam=18
\newfam\itfam
\newfam\bffam
\newfam\bfsfam
\newfam\bmitfam

\def\@mathfontinit{\count@=4
     \loop\textfont\count@=\nullfont
          \scriptfont\count@=\nullfont
          \scriptscriptfont\count@=\nullfont
          \ifnum\count@<\maxfam\advance\count@ by\@ne
     \repeat}

\def\@fontstyleinit{%
     \def\it{\err@fontnotavailable\it}%
     \def\bf{\err@fontnotavailable\bf}%
     \def\bfs{\err@bfstobf}%
     \def\bmit{\err@fontnotavailable\bmit}%
     \def\sc{\err@fontnotavailable\sc}%
     \def\sl{\err@sltoit}%
     \def\ss{\err@fontnotavailable\ss}%
     \def\tt{\err@fontnotavailable\tt}}

\def\@parameterinit#1{\rm\rp@=.1em \@getscaling{#1}%
     \let\^^\=\@doscaling\scalingskipslist
     \setbox\strutbox=\hbox{\vrule
          height.708\baselineskip depth.292\baselineskip width\z@}}

\def\@getfactor#1#2#3#4{\@@factori=#1 \@@factorii=#2
     \@@factoriii=#3 \@@factoriv=#4}

\def\@getscaling#1{\count@=#1 \advance\count@ by-\@@sizeindex\@@sizeindex=#1
     \ifnum\count@<0
          \let\@mulordiv=\divide
          \let\@divormul=\multiply
          \multiply\count@ by\m@ne
     \else\let\@mulordiv=\multiply
          \let\@divormul=\divide
     \fi
     \edef\@@scratcha{\ifcase\count@                {1}{1}{1}{1}\or
          {1}{7}{23}{3}\or     {2}{5}{3}{1}\or      {9}{89}{13}{1}\or
          {6}{25}{6}{1}\or     {8}{71}{14}{1}\or    {6}{25}{36}{5}\or
          {1}{7}{53}{4}\or     {12}{125}{108}{5}\or {3}{14}{53}{5}\or
          {6}{41}{17}{1}\or    {13}{31}{13}{2}\or   {9}{107}{71}{2}\or
          {11}{139}{124}{3}\or {1}{6}{43}{2}\or     {10}{107}{42}{1}\or
          {1}{5}{43}{2}\or     {5}{69}{65}{1}\or    {11}{97}{91}{2}\fi}%
     \expandafter\@getfactor\@@scratcha}

\def\@doscaling#1{\@mulordiv#1by\@@factori\@divormul#1by\@@factorii
     \@mulordiv#1by\@@factoriii\@divormul#1by\@@factoriv}


\newskip\headskip
\newskip\footskip

\def\typesize=#1pt{\count@=#1 \advance\count@ by-10
     \ifcase\count@
          \@setsizex\or\err@badtypesize\or
          \@setsizexii\or\err@badtypesize\or
          \@setsizexiv
     \else\err@badtypesize
     \fi}

\def\@setsizex{\getixpt
     \def\subsubscriptfonts{\vpt}%
          \def\subsubscriptsize{\vpt\@parameterinit{-8}}%
     \def\subscriptfonts{\viipt}\def\subscriptsize{\viipt\@parameterinit{-4}}%
     \def\footnotefonts{\viiipt}\def\footnotesize{\viiipt\@parameterinit{-2}}%
     \def\smallfonts{\ixpt}\def\smallsize{\ixpt\@parameterinit{-1}}%
     \def\normalfonts{\xpt}\def\normalsize{\xpt\@parameterinit{0}}%
     \def\bigfonts{\xiipt}\def\bigsize{\xiipt\@parameterinit{2}}%
     \def\Bigfonts{\xivpt}\def\Bigsize{\xivpt\@parameterinit{4}}%
     \def\biggfonts{\xviipt}\def\biggsize{\xviipt\@parameterinit{6}}%
     \def\Biggfonts{\xxipt}\def\Biggsize{\xxipt\@parameterinit{8}}%
     \def\tinyfonts{\vpt}\def\tinysize{\vpt\@parameterinit{-8}}%
     \def\HUGEFONTS{\xxvpt}\def\HUGESIZE{\xxvpt\@parameterinit{10}}%
     \normalsize\fixedskipslist}

\def\@setsizexii{\getxipt
     \def\subsubscriptfonts{\vipt}%
          \def\subsubscriptsize{\vipt\@parameterinit{-6}}%
     \def\subscriptfonts{\viiipt}%
          \def\subscriptsize{\viiipt\@parameterinit{-2}}%
     \def\footnotefonts{\xpt}\def\footnotesize{\xpt\@parameterinit{0}}%
     \def\smallfonts{\xipt}\def\smallsize{\xipt\@parameterinit{1}}%
     \def\normalfonts{\xiipt}\def\normalsize{\xiipt\@parameterinit{2}}%
     \def\bigfonts{\xivpt}\def\bigsize{\xivpt\@parameterinit{4}}%
     \def\Bigfonts{\xviipt}\def\Bigsize{\xviipt\@parameterinit{6}}%
     \def\biggfonts{\xxipt}\def\biggsize{\xxipt\@parameterinit{8}}%
     \def\Biggfonts{\xxvpt}\def\Biggsize{\xxvpt\@parameterinit{10}}%
     \def\tinyfonts{\vpt}\def\tinysize{\vpt\@parameterinit{-8}}%
     \def\HUGEFONTS{\xxvpt}\def\HUGESIZE{\xxvpt\@parameterinit{10}}%
     \normalsize\fixedskipslist}

\def\@setsizexiv{\getxiiipt
     \def\subsubscriptfonts{\viipt}%
          \def\subsubscriptsize{\viipt\@parameterinit{-4}}%
     \def\subscriptfonts{\xpt}\def\subscriptsize{\xpt\@parameterinit{0}}%
     \def\footnotefonts{\xiipt}\def\footnotesize{\xiipt\@parameterinit{2}}%
     \def\smallfonts{\xiiipt}\def\smallsize{\xiiipt\@parameterinit{3}}%
     \def\normalfonts{\xivpt}\def\normalsize{\xivpt\@parameterinit{4}}%
     \def\bigfonts{\xviipt}\def\bigsize{\xviipt\@parameterinit{6}}%
     \def\Bigfonts{\xxipt}\def\Bigsize{\xxipt\@parameterinit{8}}%
     \def\biggfonts{\xxvpt}\def\biggsize{\xxvpt\@parameterinit{10}}%
     \def\Biggfonts{\err@sizetoolarge\Biggfonts\HUGEFONTS}%
          \def\Biggsize{\err@sizetoolarge\Biggsize\HUGESIZE}%
     \def\tinyfonts{\vpt}\def\tinysize{\vpt\@parameterinit{-8}}%
     \def\HUGEFONTS{\xxvpt}\def\HUGESIZE{\xxvpt\@parameterinit{10}}%
     \normalsize\fixedskipslist}

\def\subsubscriptfonts{\vpt} \def\subsubscriptsize{\vpt\@parameterinit{-8}}
\def\subscriptfonts{\viipt}  \def\subscriptsize{\viipt\@parameterinit{-4}}
\def\footnotefonts{\viiipt}  \def\footnotesize{\viiipt\@parameterinit{-2}}
\def\smallfonts{\err@sizenotavailable\smallfonts}
                             \def\smallsize{\ixpt\@parameterinit{-1}}
\def\normalfonts{\xpt}       \def\normalsize{\xpt\@parameterinit{0}}
\def\bigfonts{\xiipt}        \def\bigsize{\xiipt\@parameterinit{2}}
\def\Bigfonts{\xivpt}        \def\Bigsize{\xivpt\@parameterinit{4}}
\def\biggfonts{\xviipt}      \def\biggsize{\xviipt\@parameterinit{6}}
\def\Biggfonts{\xxipt}       \def\Biggsize{\xxipt\@parameterinit{8}}
\def\tinyfonts{\vpt}         \def\tinysize{\vpt\@parameterinit{-8}}
\def\HUGEFONTS{\xxvpt}       \def\HUGESIZE{\xxvpt\@parameterinit{10}}

\message{document layout,}


\newtoks\everyoutput \everyoutput={}
\newdimen\depthofpage
\newcount\pagenum \pagenum=0

\newdimen\oddtopmargin  \newdimen\eventopmargin
\newdimen\oddleftmargin \newdimen\evenleftmargin
\newtoks\oddhead        \newtoks\evenhead
\newtoks\oddfoot        \newtoks\evenfoot

\def\topmargin{\afterassignment\@seteventop\oddtopmargin}
\def\leftmargin{\afterassignment\@setevenleft\oddleftmargin}
\def\head{\afterassignment\@setevenhead\oddhead}
\def\foot{\afterassignment\@setevenfoot\oddfoot}

\def\@seteventop{\eventopmargin=\oddtopmargin}
\def\@setevenleft{\evenleftmargin=\oddleftmargin}
\def\@setevenhead{\evenhead=\oddhead}
\def\@setevenfoot{\evenfoot=\oddfoot}

\def\pagenumstyle#1{\@setnumstyle\pagenum{#1}}

\newif\ifdraft
\def\draft{\drafttrue\leftmargin=.5in \overfullrule=5pt }

\def\outputstyle#1{\global\expandafter\let\expandafter
          \@outputstyle\csname#1output\endcsname
     \usename{#1setup}}

\output={\@outputstyle}

\def\normaloutput{\the\everyoutput
     \global\advance\pagenum by\@ne
     \ifodd\pagenum
          \voffset=\oddtopmargin \hoffset=\oddleftmargin
     \else\voffset=\eventopmargin \hoffset=\evenleftmargin
     \fi
     \advance\voffset by-1in  \advance\hoffset by-1in
     \count0=\pagenum
     \expandafter\shipout\pagebox
     \ifnum\outputpenalty>-\@MM\else\dosupereject\fi}

\newdimen\fullhsize
\newbox\leftpage
\newcount\leftpagenum
\newcount\outputpagenum \outputpagenum=0
\let\leftorright=L

\def\twoupoutput{\the\everyoutput
     \global\advance\pagenum by\@ne
     \if L\leftorright
          \global\setbox\leftpage=\leftline{\pagebox}%
          \global\leftpagenum=\pagenum
          \global\let\leftorright=R%
     \else\global\advance\outputpagenum by\@ne
          \ifodd\outputpagenum
               \voffset=\oddtopmargin \hoffset=\oddleftmargin
          \else\voffset=\eventopmargin \hoffset=\evenleftmargin
          \fi
          \advance\voffset by-1in  \advance\hoffset by-1in
          \count0=\leftpagenum \count1=\pagenum
          \shipout\vbox{\hbox to\fullhsize
               {\box\leftpage\hfil\leftline{\pagebox}}}%
          \global\let\leftorright=L%
     \fi
     \ifnum\outputpenalty>-\@MM
     \else\dosupereject
          \if R\leftorright
               \globaldefs=\@ne\head={\hfil}\foot={\hfil}\globaldefs=\z@
               \null\newpage
          \fi
     \fi}

\def\pagebox{\vbox{\makeheadline\pagebody\makefootline}}

\def\makeheadline{%
     \vbox to\z@{\baselinestretch=\@m
          \vskip\topskip\vskip-.708\baselineskip\vskip-\headskip
          \line{\vbox to\ht\strutbox{}%
               \ifodd\pagenum\the\oddhead\else\the\evenhead\fi}%
          \vss}%
     \nointerlineskip}

\def\pagebody{\vbox to\vsize{%
     \boxmaxdepth\maxdepth
     \ifvoid\topins\else\unvbox\topins\fi
     \depthofpage=\dp255
     \unvbox255
     \ifraggedbottom\kern-\depthofpage\vfil\fi
     \ifvoid\footins
     \else\vskip\skip\footins
          \footnoterule
          \unvbox\footins
          \vskip-\footnoteskip
     \fi}}

\def\makefootline{\baselineskip=\footskip
     \line{\ifodd\pagenum\the\oddfoot\else\the\evenfoot\fi}}


\newskip\abovechapterskip
\newskip\belowchapterskip
\newskip\abovesectionskip
\newskip\belowsectionskip
\newskip\abovesubsectionskip
\newskip\belowsubsectionskip

\def\chapterstyle#1{\global\expandafter\let\expandafter\@chapterstyle
     \csname#1text\endcsname}
\def\sectionstyle#1{\global\expandafter\let\expandafter\@sectionstyle
     \csname#1text\endcsname}
\def\subsectionstyle#1{\global\expandafter\let\expandafter\@subsectionstyle
     \csname#1text\endcsname}

\def\chapter#1{%
     \ifdim\lastskip=17sp \else\chapterbreak\vskip\abovechapterskip\fi
     \@chapterstyle{\ifblank\chapternumstyle\then
          \else\newchapternum=\next\chapternumformat\ \fi#1}%
     \nobreak\vskip\belowchapterskip\vskip17sp }

\def\section#1{%
     \ifdim\lastskip=17sp \else\sectionbreak\vskip\abovesectionskip\fi
     \@sectionstyle{\ifblank\sectionnumstyle\then
          \else\newsectionnum=\next\sectionnumformat\ \fi#1}%
     \nobreak\vskip\belowsectionskip\vskip17sp }

\def\subsection#1{%
     \ifdim\lastskip=17sp \else\subsectionbreak\vskip\abovesubsectionskip\fi
     \@subsectionstyle{\ifblank\subsectionnumstyle\then
          \else\newsubsectionnum=\next\subsectionnumformat\ \fi#1}%
     \nobreak\vskip\belowsubsectionskip\vskip17sp }


\let\TeXunderline=\underline
\let\TeXoverline=\overline
\def\underline#1{\relax\ifmmode\TeXunderline{#1}\else
     $\TeXunderline{\hbox{#1}}$\fi}
\def\overline#1{\relax\ifmmode\TeXoverline{#1}\else
     $\TeXoverline{\hbox{#1}}$\fi}

\def\baselinestretch{\afterassignment\@baselinestretch\count@}
\def\@baselinestretch{\baselineskip=\normalbaselineskip
     \divide\baselineskip by\@m\baselineskip=\count@\baselineskip
     \setbox\strutbox=\hbox{\vrule
          height.708\baselineskip depth.292\baselineskip width\z@}%
     \bigskipamount=\the\baselineskip
          plus.25\baselineskip minus.25\baselineskip
     \medskipamount=.5\baselineskip
          plus.125\baselineskip minus.125\baselineskip
     \smallskipamount=.25\baselineskip
          plus.0625\baselineskip minus.0625\baselineskip}

\def\\{\ifhmode\ifnum\lastpenalty=-\@M\else\hfil\penalty-\@M\fi\fi
     \ignorespaces}
\def\newpage{\vfil\break}

\def\lefttext#1{\par{\@text\leftskip=\z@\rightskip=\centering
     \noindent#1\par}}
\def\righttext#1{\par{\@text\leftskip=\centering\rightskip=\z@
     \noindent#1\par}}
\def\centertext#1{\par{\@text\leftskip=\centering\rightskip=\centering
     \noindent#1\par}}
\def\@text{\parindent=\z@ \parfillskip=\z@ \everypar={}%
     \spaceskip=.3333em \xspaceskip=.5em
     \def\\{\ifhmode\ifnum\lastpenalty=-\@M\else\penalty-\@M\fi\fi
          \ignorespaces}}

\def\beginleft{\par\@text\leftskip=\z@ \rightskip=\centering}
     
\def\beginright{\par\@text\leftskip=\centering\rightskip=\z@ }
     
\def\begincenter{\par\@text\leftskip=\centering\rightskip=\centering}

\def\beginnarrow{\defaultoption[\parindent]\@beginnarrow}
\def\@beginnarrow[#1]{\par\advance\leftskip by#1\advance\rightskip by#1}

\begingroup
\catcode`\[=1 \catcode`\{=11 \gdef\beginignore[\endgroup\bgroup
     \catcode`\e=0 \catcode`\\=12 \catcode`\{=11 \catcode`\f=12 \let\or=\relax
     \let\nd{ignor=\fi \let\}=\egroup
     \iffalse}
\endgroup

\long\def\marginnote#1{\leavevmode
     \edef\@marginsf{\spacefactor=\the\spacefactor\relax}%
     \ifdraft\strut\vadjust{%
          \hbox to\z@{\hskip\hsize\hskip.1in
               \vbox to\z@{\vskip-\dp\strutbox
                    \marginnoteformat
                    \vskip-\ht\strutbox
                    \noindent\strut#1\par
                    \vss}%
               \hss}}%
     \fi
     \@marginsf}


\newtoks\everybye \everybye={\par\vfil}
\outer\def\bye{\the\everybye
     \footnotecheck
     \prelabelcheck
     \streamcheck
     \supereject
     \TeXend}

\message{footnotes,}

\newcount\footnotenum \footnotenum=0
\newskip\footnoteskip
\let\@footnotelist=\empty

\def\footnotenumstyle#1{\@setnumstyle\footnotenum{#1}%
     \useafter\ifx{@footnotenumstyle}\symbols
          \global\let\@footup=\empty
     \else\global\let\@footup=\markup
     \fi}

\def\footnote{\footnotecheck\defaultoption[]\@footnote}
\def\@footnote[#1]{\@footnotemark[#1]\@footnotetext}

\def\footnotemark{\defaultoption[]\@footnotemark}
\def\@footnotemark[#1]{\let\@footsf=\empty
     \ifhmode\edef\@footsf{\spacefactor=\the\spacefactor\relax}\/\fi
     \ifnoarg#1\then
          \global\advance\footnotenum by\@ne
          \@footup{\footnotenumformat}%
          \edef\@@foota{\footnotenum=\the\footnotenum\relax}%
          \expandafter\additemR\expandafter\@footup\expandafter
               {\@@foota\footnotenumformat}\to\@footnotelist
          \global\let\@footnotelist=\@footnotelist
     \else\markup{#1}%
          \additemR\markup{#1}\to\@footnotelist
          \global\let\@footnotelist=\@footnotelist
     \fi
     \@footsf}

\def\footnotetext{%
     \ifx\@footnotelist\empty\err@extrafootnotetext\else\@footnotetext\fi}
\def\@footnotetext{%
     \getitemL\@footnotelist\to\@@foota
     \global\let\@footnotelist=\@footnotelist
     \insert\footins\bgroup
     \footnoteformat
     \splittopskip=\ht\strutbox\splitmaxdepth=\dp\strutbox
     \interlinepenalty=\interfootnotelinepenalty\floatingpenalty=\@MM
     \noindent\llap{\@@foota}\strut
     \bgroup\aftergroup\@footnoteend
     \let\@@scratcha=}
\def\@footnoteend{\strut\par\vskip\footnoteskip\egroup}

\def\footnoterule{\normalfonts
     \kern-.3em \hrule width2in height.04em \kern .26em }

\def\footnotecheck{%
     \ifx\@footnotelist\empty
     \else\err@extrafootnotemark
          \global\let\@footnotelist=\empty
     \fi}

\message{labels,}

\let\@@labeldef=\xdef
\newif\if@labelfile
\newwrite\@labelfile
\let\@prelabellist=\empty

\def\label#1#2{\trim#1\to\@@labarg\edef\@@labtext{#2}%
     \edef\@@labname{lab@\@@labarg}%
     \useafter\ifundefined\@@labname\then\else\@yeslab\fi
     \useafter\@@labeldef\@@labname{#2}%
     \ifstreaming
          \expandafter\toks@\expandafter\expandafter\expandafter
               {\csname\@@labname\endcsname}%
          \immediate\write\streamout{\noexpand\label{\@@labarg}{\the\toks@}}%
     \fi}
\def\@yeslab{%
     \useafter\ifundefined{if\@@labname}\then
          \err@labelredef\@@labarg
     \else\useif{if\@@labname}\then
               \err@labelredef\@@labarg
          \else\global\usename{\@@labname true}%
               \useafter\ifundefined{pre\@@labname}\then
               \else\useafter\ifx{pre\@@labname}\@@labtext
                    \else\err@badlabelmatch\@@labarg
                    \fi
               \fi
               \if@labelfile
               \else\global\@labelfiletrue
                    \immediate\write\sixt@@n{--> Creating file \jobname.lab}%
                    \immediate\openout\@labelfile=\jobname.lab
               \fi
               \immediate\write\@labelfile
                    {\noexpand\prelabel{\@@labarg}{\@@labtext}}%
          \fi
     \fi}

\def\putlab#1{\trim#1\to\@@labarg\edef\@@labname{lab@\@@labarg}%
     \useafter\ifundefined\@@labname\then\@nolab\else\usename\@@labname\fi}
\def\@nolab{%
     \useafter\ifundefined{pre\@@labname}\then
          \undefinedlabelformat
          \err@needlabel\@@labarg
          \useafter\xdef\@@labname{\undefinedlabelformat}%
     \else\usename{pre\@@labname}%
          \useafter\xdef\@@labname{\usename{pre\@@labname}}%
     \fi
     \useafter\newif{if\@@labname}%
     \expandafter\additemR\@@labarg\to\@prelabellist}

\def\prelabel#1{\useafter\gdef{prelab@#1}}

\def\ifundefinedlabel#1\then{%
     \expandafter\ifx\csname lab@#1\endcsname\relax}
\def\useiflab#1\then{\csname iflab@#1\endcsname}

\def\prelabelcheck{{%
     \def\^^\##1{\useiflab{##1}\then\else\err@undefinedlabel{##1}\fi}%
     \@prelabellist}}

\message{equation numbering,}

\newcount\chapternum
\newcount\sectionnum
\newcount\subsectionnum
\newcount\equationnum
\newcount\subequationnum
\newcount\figurenum
\newcount\subfigurenum
\newcount\tablenum
\newcount\subtablenum

\newif\if@subeqncount
\newif\if@subfigcount
\newif\if@subtblcount

\def\newchapternum{\newsectionnum=\z@\@resetnum\chapternum}
\def\newsectionnum{\newsubsectionnum=\z@\@resetnum\sectionnum}
\def\newsubsectionnum{\newequationnum=\z@\newfigurenum=\z@\newtablenum=\z@
     \@resetnum\subsectionnum}
\def\newequationnum{\newsubequationnum=\z@\@resetnum\equationnum}
\def\newsubequationnum{\@resetnum\subequationnum}
\def\newfigurenum{\newsubfigurenum=\z@\@resetnum\figurenum}
\def\newsubfigurenum{\@resetnum\subfigurenum}
\def\newtablenum{\newsubtablenum=\z@\@resetnum\tablenum}
\def\newsubtablenum{\@resetnum\subtablenum}

\def\@resetnum#1{\global\advance#1by1 \edef\next{\the#1\relax}\global#1}

\newchapternum=0

\def\chapternumstyle#1{\@setnumstyle\chapternum{#1}}
\def\sectionnumstyle#1{\@setnumstyle\sectionnum{#1}}
\def\subsectionnumstyle#1{\@setnumstyle\subsectionnum{#1}}
\def\equationnumstyle#1{\@setnumstyle\equationnum{#1}}
\def\subequationnumstyle#1{\@setnumstyle\subequationnum{#1}%
     \ifblank\subequationnumstyle\then\global\@subeqncountfalse\fi
     \ignorespaces}
\def\figurenumstyle#1{\@setnumstyle\figurenum{#1}}
\def\subfigurenumstyle#1{\@setnumstyle\subfigurenum{#1}%
     \ifblank\subfigurenumstyle\then\global\@subfigcountfalse\fi
     \ignorespaces}
\def\tablenumstyle#1{\@setnumstyle\tablenum{#1}}
\def\subtablenumstyle#1{\@setnumstyle\subtablenum{#1}%
     \ifblank\subtablenumstyle\then\global\@subtblcountfalse\fi
     \ignorespaces}

\def\eqnlabel#1{%
     \if@subeqncount
          \newsubequationnum=\next
     \else\newequationnum=\next
          \ifblank\subequationnumstyle\then
          \else\global\@subeqncounttrue
               \newsubequationnum=\@ne
          \fi
     \fi
     \label{#1}{\puteqnformat}(\puteqn{#1})%
     \ifdraft\rlap{\hskip.1in{\tt#1}}\fi}

\let\puteqn=\putlab

\def\equation#1#2{\useafter\gdef{eqn@#1}{#2\eqno\eqnlabel{#1}}}
\def\Equation#1{\useafter\gdef{eqn@#1}}

\def\putequation#1{\useafter\ifundefined{eqn@#1}\then
     \err@undefinedeqn{#1}\else\usename{eqn@#1}\fi}

\def\eqnseriesstyle#1{\gdef\@eqnseriesstyle{#1}}
\def\begineqnseries{\subequationnumstyle{\@eqnseriesstyle}%
     \defaultoption[]\@begineqnseries}
\def\@begineqnseries[#1]{\edef\@@eqnname{#1}}
\def\endeqnseries{\subequationnumstyle{blank}%
     \expandafter\ifnoarg\@@eqnname\then
     \else\label\@@eqnname{\puteqnformat}%
     \fi
     \aftergroup\ignorespaces}

\def\figlabel#1{%
     \if@subfigcount
          \newsubfigurenum=\next
     \else\newfigurenum=\next
          \ifblank\subfigurenumstyle\then
          \else\global\@subfigcounttrue
               \newsubfigurenum=\@ne
          \fi
     \fi
     \label{#1}{\putfigformat}\putfig{#1}%
     {\def\marginnoteformat{\tt}\marginnote{#1}}}

\let\putfig=\putlab

\def\figseriesstyle#1{\gdef\@figseriesstyle{#1}}
\def\beginfigseries{\subfigurenumstyle{\@figseriesstyle}%
     \defaultoption[]\@beginfigseries}
\def\@beginfigseries[#1]{\edef\@@figname{#1}}
\def\endfigseries{\subfigurenumstyle{blank}%
     \expandafter\ifnoarg\@@figname\then
     \else\label\@@figname{\putfigformat}%
     \fi
     \aftergroup\ignorespaces}

\def\tbllabel#1{%
     \if@subtblcount
          \newsubtablenum=\next
     \else\newtablenum=\next
          \ifblank\subtablenumstyle\then
          \else\global\@subtblcounttrue
               \newsubtablenum=\@ne
          \fi
     \fi
     \label{#1}{\puttblformat}\puttbl{#1}%
     {\def\marginnoteformat{\tt}\marginnote{#1}}}

\let\puttbl=\putlab

\def\tblseriesstyle#1{\gdef\@tblseriesstyle{#1}}
\def\begintblseries{\subtablenumstyle{\@tblseriesstyle}%
     \defaultoption[]\@begintblseries}
\def\@begintblseries[#1]{\edef\@@tblname{#1}}
\def\endtblseries{\subtablenumstyle{blank}%
     \expandafter\ifnoarg\@@tblname\then
     \else\label\@@tblname{\puttblformat}%
     \fi
     \aftergroup\ignorespaces}

\message{reference numbering,}

\newcount\referencenum \referencenum=0
\newcount\@@prerefcount \@@prerefcount=0
\newcount\@@thisref
\newcount\@@lastref
\newcount\@@loopref
\newcount\@@refseq
\newdimen\refnumindent
\let\@undefreflist=\empty

\def\referencenumstyle#1{\@setnumstyle\referencenum{#1}}

\def\referencestyle#1{\usename{@ref#1}}

\def\@refsequential{%
     \gdef\@refpredef##1{\global\advance\referencenum by\@ne
          \let\^^\=0\label{##1}{\^^\{\the\referencenum}}%
          \useafter\gdef{ref@\the\referencenum}{{##1}{\undefinedlabelformat}}}%
     \gdef\@reference##1##2{%
          \ifundefinedlabel##1\then
          \else\def\^^\####1{\global\@@thisref=####1\relax}\putlab{##1}%
               \useafter\gdef{ref@\the\@@thisref}{{##1}{##2}}%
          \fi}%
     \gdef\endputreferences{%
          \loop\ifnum\@@loopref<\referencenum
                    \advance\@@loopref by\@ne
                    \expandafter\expandafter\expandafter\@printreference
                         \csname ref@\the\@@loopref\endcsname
          \repeat
          \par}}

\def\@refpreordered{%
     \gdef\@refpredef##1{\global\advance\referencenum by\@ne
          \additemR##1\to\@undefreflist}%
     \gdef\@reference##1##2{%
          \ifundefinedlabel##1\then
          \else\global\advance\@@loopref by\@ne
               {\let\^^\=0\label{##1}{\^^\{\the\@@loopref}}}%
               \@printreference{##1}{##2}%
          \fi}
     \gdef\endputreferences{%
          \def\^^\####1{\useiflab{####1}\then
               \else\reference{####1}{\undefinedlabelformat}\fi}%
          \@undefreflist
          \par}}

\def\beginprereferences{\par
     \def\reference##1##2{\global\advance\referencenum by1\@ne
          \let\^^\=0\label{##1}{\^^\{\the\referencenum}}%
          \useafter\gdef{ref@\the\referencenum}{{##1}{##2}}}}
\def\endprereferences{\global\@@prerefcount=\the\referencenum\par}

\def\beginputreferences{\par
     \refnumindent=\z@\@@loopref=\z@
     \loop\ifnum\@@loopref<\referencenum
               \advance\@@loopref by\@ne
               \setbox\z@=\hbox{\referencenum=\@@loopref
                    \referencenumformat\enskip}%
               \ifdim\wd\z@>\refnumindent\refnumindent=\wd\z@\fi
     \repeat
     \putreferenceformat
     \@@loopref=\z@
     \loop\ifnum\@@loopref<\@@prerefcount
               \advance\@@loopref by\@ne
               \expandafter\expandafter\expandafter\@printreference
                    \csname ref@\the\@@loopref\endcsname
     \repeat
     \let\reference=\@reference}

\def\@printreference#1#2{\ifx#2\undefinedlabelformat\err@undefinedref{#1}\fi
     \noindent\ifdraft\rlap{\hskip\hsize\hskip.1in \tt#1}\fi
     \llap{\referencenum=\@@loopref\referencenumformat\enskip}#2\par}

\def\reference#1#2{{\par\refnumindent=\z@\putreferenceformat\noindent#2\par}}

\def\putref#1{\trim#1\to\@@refarg
     \expandafter\ifnoarg\@@refarg\then
          \toks@={\relax}%
     \else\@@lastref=-\@m\def\@@refsep{}\def\@more{\@nextref}%
          \toks@={\@nextref#1,,}%
     \fi\the\toks@}
\def\@nextref#1,{\trim#1\to\@@refarg
     \expandafter\ifnoarg\@@refarg\then
          \let\@more=\relax
     \else\ifundefinedlabel\@@refarg\then
               \expandafter\@refpredef\expandafter{\@@refarg}%
          \fi
          \def\^^\##1{\global\@@thisref=##1\relax}%
          \global\@@thisref=\m@ne
          \setbox\z@=\hbox{\putlab\@@refarg}%
     \fi
     \advance\@@lastref by\@ne
     \ifnum\@@lastref=\@@thisref\advance\@@refseq by\@ne\else\@@refseq=\@ne\fi
     \ifnum\@@lastref<\z@
     \else\ifnum\@@refseq<\thr@@
               \@@refsep\def\@@refsep{,}%
               \ifnum\@@lastref>\z@
                    \advance\@@lastref by\m@ne
                    {\referencenum=\@@lastref\putrefformat}%
               \else\undefinedlabelformat
               \fi
          \else\def\@@refsep{--}%
          \fi
     \fi
     \@@lastref=\@@thisref
     \@more}

\message{streaming,}

\newif\ifstreaming

\def\streamto{\defaultoption[\jobname]\@streamto}
\def\@streamto[#1]{\global\streamingtrue
     \immediate\write\sixt@@n{--> Streaming to #1.str}%
     \newwrite\streamout\immediate\openout\streamout=#1.str }

\def\streamfrom{\defaultoption[\jobname]\@streamfrom}
\def\@streamfrom[#1]{\newread\streamin\openin\streamin=#1.str
     \ifeof\streamin
          \expandafter\err@nostream\expandafter{#1.str}%
     \else\immediate\write\sixt@@n{--> Streaming from #1.str}%
          \let\@@labeldef=\gdef
          \ifstreaming
               \edef\@elc{\endlinechar=\the\endlinechar}%
               \endlinechar=\m@ne
               \loop\read\streamin to\@@scratcha
                    \ifeof\streamin
                         \streamingfalse
                    \else\toks@=\expandafter{\@@scratcha}%
                         \immediate\write\streamout{\the\toks@}%
                    \fi
                    \ifstreaming
               \repeat
               \@elc
               \input #1.str
               \streamingtrue
          \else\input #1.str
          \fi
          \let\@@labeldef=\xdef
     \fi}

\def\streamcheck{\ifstreaming
     \immediate\write\streamout{\pagenum=\the\pagenum}%
     \immediate\write\streamout{\footnotenum=\the\footnotenum}%
     \immediate\write\streamout{\referencenum=\the\referencenum}%
     \immediate\write\streamout{\chapternum=\the\chapternum}%
     \immediate\write\streamout{\sectionnum=\the\sectionnum}%
     \immediate\write\streamout{\subsectionnum=\the\subsectionnum}%
     \immediate\write\streamout{\equationnum=\the\equationnum}%
     \immediate\write\streamout{\subequationnum=\the\subequationnum}%
     \immediate\write\streamout{\figurenum=\the\figurenum}%
     \immediate\write\streamout{\subfigurenum=\the\subfigurenum}%
     \immediate\write\streamout{\tablenum=\the\tablenum}%
     \immediate\write\streamout{\subtablenum=\the\subtablenum}%
     \immediate\closeout\streamout
     \fi}


\def\err@badtypesize{%
     \errhelp={The limited availability of certain fonts requires^^J%
          that the base type size be 10pt, 12pt, or 14pt.^^J}%
     \errmessage{--> Illegal base type size}}

\def\err@badsizechange{\immediate\write\sixt@@n
     {--> Size change not allowed in math mode, ignored}}

\def\err@sizetoolarge#1{\immediate\write\sixt@@n
     {--> \noexpand#1 too big, substituting HUGE}}

\def\err@sizenotavailable#1{\immediate\write\sixt@@n
     {--> Size not available, \noexpand#1 ignored}}

\def\err@fontnotavailable#1{\immediate\write\sixt@@n
     {--> Font not available, \noexpand#1 ignored}}

\def\err@sltoit{\immediate\write\sixt@@n
     {--> Style \noexpand\sl not available, substituting \noexpand\it}%
     \it}

\def\err@bfstobf{\immediate\write\sixt@@n
     {--> Style \noexpand\bfs not available, substituting \noexpand\bf}%
     \bf}

\def\err@badgroup#1#2{%
     \errhelp={The block you have just tried to close was not the one^^J%
          most recently opened.^^J}%
     \errmessage{--> \noexpand\end{#1} doesn't match \noexpand\begin{#2}}}

\def\err@badcountervalue#1{\immediate\write\sixt@@n
     {--> Counter (#1) out of bounds}}

\def\err@extrafootnotemark{\immediate\write\sixt@@n
     {--> \noexpand\footnotemark command
          has no corresponding \noexpand\footnotetext}}

\def\err@extrafootnotetext{%
     \errhelp{You have given a \noexpand\footnotetext command without first
          specifying^^Ja \noexpand\footnotemark.^^J}%
     \errmessage{--> \noexpand\footnotetext command has no corresponding
          \noexpand\footnotemark}}

\def\err@labelredef#1{\immediate\write\sixt@@n
     {--> Label "#1" redefined}}

\def\err@badlabelmatch#1{\immediate\write\sixt@@n
     {--> Definition of label "#1" doesn't match value in \jobname.lab}}

\def\err@needlabel#1{\immediate\write\sixt@@n
     {--> Label "#1" cited before its definition}}

\def\err@undefinedlabel#1{\immediate\write\sixt@@n
     {--> Label "#1" cited but never defined}}

\def\err@undefinedeqn#1{\immediate\write\sixt@@n
     {--> Equation "#1" not defined}}

\def\err@undefinedref#1{\immediate\write\sixt@@n
     {--> Reference "#1" not defined}}

\def\err@nostream#1{%
     \errhelp={You have tried to input a stream file that doesn't exist.^^J}%
     \errmessage{--> Stream file #1 not found}}

\message{jyTeX initialization}

\everyjob{\immediate\write16{--> jyTeX version \fmtversion}%
     \edef\@@jobname{\jobname}%
     \edef\jobname{\@@jobname}%
     \settime
     \openin0=\jobname.lab
     \ifeof0
     \else\closein0
          \immediate\write16{--> Getting labels from file \jobname.lab}%
          \input\jobname.lab
     \fi}


\def\fixedskipslist{%
     \^^\{\topskip}%
     \^^\{\splittopskip}%
     \^^\{\maxdepth}%
     \^^\{\skip\topins}%
     \^^\{\skip\footins}%
     \^^\{\headskip}%
     \^^\{\footskip}}

\def\scalingskipslist{%
     \^^\{\p@renwd}%
     \^^\{\delimitershortfall}%
     \^^\{\nulldelimiterspace}%
     \^^\{\scriptspace}%
     \^^\{\jot}%
     \^^\{\normalbaselineskip}%
     \^^\{\normallineskip}%
     \^^\{\normallineskiplimit}%
     \^^\{\baselineskip}%
     \^^\{\lineskip}%
     \^^\{\lineskiplimit}%
     \^^\{\bigskipamount}%
     \^^\{\medskipamount}%
     \^^\{\smallskipamount}%
     \^^\{\parskip}%
     \^^\{\parindent}%
     \^^\{\abovedisplayskip}%
     \^^\{\belowdisplayskip}%
     \^^\{\abovedisplayshortskip}%
     \^^\{\belowdisplayshortskip}%
     \^^\{\abovechapterskip}%
     \^^\{\belowchapterskip}%
     \^^\{\abovesectionskip}%
     \^^\{\belowsectionskip}%
     \^^\{\abovesubsectionskip}%
     \^^\{\belowsubsectionskip}}


\def\twoupsetup{
     \topmargin=.75in
     \leftmargin=.5in
     \vsize=6.9in
     \hsize=4.75in
     \fullhsize=10in
     \let\draft=\relax}

\outputstyle{normal}                             

\def\marginnoteformat{\subscriptsize             
     \hsize=1in \baselinestretch=1000 \everypar={}%
     \tolerance=5000 \hbadness=5000 \parskip=0pt \parindent=0pt
     \leftskip=0pt \rightskip=0pt \raggedright}

\head={\ifdraft\normalfonts\it\hfil DRAFT\hfil   
     \llap{\number\day\ \monthword\month\ \militarytime}\else\hfil\fi}
\foot={\hfil\normalfonts\numstyle\pagenum\hfil}  

\normalbaselineskip=12pt                         
\normallineskip=0pt                              
\normallineskiplimit=0pt                         
\normalbaselines                                 

\topskip=.85\baselineskip \splittopskip=\topskip \headskip=2\baselineskip
\footskip=\headskip

\pagenumstyle{arabic}                            

\parskip=0pt                                     
\parindent=20pt                                  

\baselinestretch=1000                            


\chapterstyle{left}                              
\chapternumstyle{blank}                          
\def\chapterbreak{\newpage}                      
\abovechapterskip=0pt                            
\belowchapterskip=1.5\baselineskip               
     plus.38\baselineskip minus.38\baselineskip
\def\chapternumformat{\numstyle\chapternum.}     

\sectionstyle{left}                              
\sectionnumstyle{blank}                          
\def\sectionbreak{\vskip0pt plus4\baselineskip\penalty-100
     \vskip0pt plus-4\baselineskip}              
\abovesectionskip=1.5\baselineskip               
     plus.38\baselineskip minus.38\baselineskip
\belowsectionskip=\the\baselineskip              
     plus.25\baselineskip minus.25\baselineskip
\def\sectionnumformat{
     \ifblank\chapternumstyle\then\else\numstyle\chapternum.\fi
     \numstyle\sectionnum.}

\subsectionstyle{left}                           
\subsectionnumstyle{blank}                       
\def\subsectionbreak{\vskip0pt plus4\baselineskip\penalty-100
     \vskip0pt plus-4\baselineskip}              
\abovesubsectionskip=\the\baselineskip           
     plus.25\baselineskip minus.25\baselineskip
\belowsubsectionskip=.75\baselineskip            
     plus.19\baselineskip minus.19\baselineskip
\def\subsectionnumformat{
     \ifblank\chapternumstyle\then\else\numstyle\chapternum.\fi
     \ifblank\sectionnumstyle\then\else\numstyle\sectionnum.\fi
     \numstyle\subsectionnum.}


\footnotenumstyle{symbols}                       
\footnoteskip=0pt                                
\def\footnotenumformat{\numstyle\footnotenum}    
\def\footnoteformat{\footnotesize                
     \everypar={}\parskip=0pt \parfillskip=0pt plus1fil
     \leftskip=1em \rightskip=0pt
     \spaceskip=0pt \xspaceskip=0pt
     \def\\{\ifhmode\ifnum\lastpenalty=-10000
          \else\hfil\penalty-10000 \fi\fi\ignorespaces}}


\def\undefinedlabelformat{$\bullet$}             


\equationnumstyle{arabic}                        
\subequationnumstyle{blank}                      
\figurenumstyle{arabic}                          
\subfigurenumstyle{blank}                        
\tablenumstyle{arabic}                           
\subtablenumstyle{blank}                         

\eqnseriesstyle{alphabetic}                      
\figseriesstyle{alphabetic}                      
\tblseriesstyle{alphabetic}                      

\def\puteqnformat{\hbox{
     \ifblank\chapternumstyle\then\else\numstyle\chapternum.\fi
     \ifblank\sectionnumstyle\then\else\numstyle\sectionnum.\fi
     \ifblank\subsectionnumstyle\then\else\numstyle\subsectionnum.\fi
     \numstyle\equationnum
     \numstyle\subequationnum}}
\def\putfigformat{\hbox{
     \ifblank\chapternumstyle\then\else\numstyle\chapternum.\fi
     \ifblank\sectionnumstyle\then\else\numstyle\sectionnum.\fi
     \ifblank\subsectionnumstyle\then\else\numstyle\subsectionnum.\fi
     \numstyle\figurenum
     \numstyle\subfigurenum}}
\def\puttblformat{\hbox{
     \ifblank\chapternumstyle\then\else\numstyle\chapternum.\fi
     \ifblank\sectionnumstyle\then\else\numstyle\sectionnum.\fi
     \ifblank\subsectionnumstyle\then\else\numstyle\subsectionnum.\fi
     \numstyle\tablenum
     \numstyle\subtablenum}}


\referencestyle{sequential}                      
\referencenumstyle{arabic}                       
\def\putrefformat{\numstyle\referencenum}        
\def\referencenumformat{\numstyle\referencenum.} 
\def\putreferenceformat{
     \everypar={\hangindent=1em \hangafter=1 }%
     \def\\{\hfil\break\null\hskip-1em \ignorespaces}%
     \leftskip=\refnumindent\parindent=0pt \interlinepenalty=1000 }


\normalsize


\def\fmtversion{2.6M (June 1992)}

\catcode`\@=12

\typesize=10pt \magnification=1200 \baselineskip17truept
\footnotenumstyle{arabic} \hsize=6truein\vsize=8.5truein
\input epsf
\sectionnumstyle{blank}
\chapternumstyle{blank}
\chapternum=1
\sectionnum=1
\pagenum=0

\def\begintitle{\pagenumstyle{blank}\parindent=0pt
\begin{narrow}[0.4in]}
\def\endtitle{\end{narrow}\newpage\pagenumstyle{arabic}}


\def\beginexercise{\vskip 20truept\parindent=0pt\begin{narrow}[10
truept]}
\def\endexercise{\vskip 10truept\end{narrow}}


\def\eql#1{\eqno\eqnlabel{#1}}
\def\ref{\reference}
\def\peq{\puteqn}
\def\pref{\putref}

\def\mgn{\marginnote}
\def\bex{\begin{exercise}}
\def\eex{\end{exercise}}




\def\StretchRtArr#1{{\count255=0\loop\relbar\joinrel\advance\count255 by1
\ifnum\count255<#1\repeat\rightarrow}}
\def\StretchLtArr#1{\,{\leftarrow\!\!\count255=0\loop\relbar
\joinrel\advance\count255 by1\ifnum\count255<#1\repeat}}

\def\StretchLRtArr#1{\,{\leftarrow\!\!\count255=0\loop\relbar\joinrel\advance
\count255 by1\ifnum\count255<#1\repeat\rightarrow\,\,}}

\def\mbox#1{{\leavevmode\hbox{#1}}}

\def\hspace#1{{\phantom{\mbox#1}}}



\def\Ga{\Gamma}

\def\Up{\Upsilon}

\def\ze{\zeta}

\def\caC{{\cal C}}

\def\caS{{\cal S}}

\def\det{{\rm det\,}}

\def\sc{{\rm sc }}

\def\zf{$\zeta$--function}
\def\zfs{$\zeta$--functions}


\def\frac#1/#2{\leavevmode\kern.1em
\raise.5ex\hbox{\the\scriptfont0 #1}\kern-.1em/\kern-.15em
\lower.25ex\hbox{\the\scriptfont0 #2}}
\def\sfrac#1/#2{\leavevmode\kern.1em
\raise.5ex\hbox{\the\scriptscriptfont0 #1}\kern-.1em/\kern-.15em
\lower.25ex\hbox{\the\scriptscriptfont0 #2}}

\def\gtorder{\mathrel{\raise.3ex\hbox{$>$}\mkern-14mu
             \lower0.6ex\hbox{$\sim$}}}
\def\ltorder{\mathrel{\raise.3ex\hbox{$<$}\mkern-14mu
             \lower0.6ex\hbox{$\sim$}}}

\def\semidirprod{\rlap{\ss C}\raise1pt\hbox{$\mkern.75mu\times$}}
\def\for{\lower6pt\hbox{$\Big|$}}
\def\fish{\kern-.25em{\phantom{abcde}\over \phantom{abcde}}\kern-.25em}


\def\boxit#1{\vbox{\hrule\hbox{\vrule\kern3pt
        \vbox{\kern3pt#1\kern3pt}\kern3pt\vrule}\hrule}}
\def\dalemb#1#2{{\vbox{\hrule height .#2pt
        \hbox{\vrule width.#2pt height#1pt \kern#1pt \vrule
                width.#2pt} \hrule height.#2pt}}}
\def\square{\mathord{\dalemb{5.9}{6}\hbox{\hskip1pt}}}

\def\frac#1#2{{{#1}\over{#2}}}

\def\noin{\noindent}

\def\comb#1#2{{\left(#1\atop#2\right)}}

\def\nsl{\nabla\!\!\!\! / }

\def\cosech{{\rm cosech\,}}

\def\eg{{\it e.g.}}
\def\ie{{\it i.e. }}



\def\wt{\widetilde}

\def\3j#1#2#3#4#5#6{\left\lgroup\matrix{#1&#2&#3\cr#4&#5&#6\cr}
\right\rgroup}

\def\m?{\mgn{?}}

\def\beq{\begin{eqnarray}}
\def\eeq{\end{eqnarray}}


\def\cmp#1#2#3{{\it Comm. Math. Phys.} {\bf {#1}} ({#2}) #3}
\def\cqg#1#2#3{{\it Class. Quant. Grav.} {\bf {#1}} ({#2}) #3}

\def\jgp#1#2#3{{\it J. Geom. and Phys.} {\bf {#1}} ({#2}) #3}

\def\jpa#1#2#3{{\it J. Phys.} {\bf A{#1}} ({#2}) #3}

\def\np#1#2#3{{\it Nucl. Phys.} {\bf B{#1}} ({#2}) #3}

\def\pl#1#2#3{{\it Phys. Lett.} {\bf {#1}} ({#2}) #3}

\def\prp#1#2#3{{\it Phys. Rep.} {\bf {#1}} ({#2}) #3}

\def\prD#1#2#3{{\it Phys. Rev.} {\bf D{#1}} ({#2}) #3}

\def\dmj#1#2#3{{\it Duke Math. J.} {\bf {#1}} ({#2}) #3}

\def\jpamt#1#2#3{{\it J. Phys.A:Math.Theor.} {\bf{#1}} ({#2}) #3}

\def\ma#1#2#3{{\it Math. Ann.} {\bf {#1}} ({#2}) #3}

\begin{title}
\vglue 0.5truein
\vskip15truept
\centertext {\Bigfonts \bf A technical note on the calculation of} \vskip7truept
\vskip10truept\centertext{\Bigfonts \bf GJMS (Rac and Di) operator determinants }
 \vskip7truept
\vskip10truept\centertext{\Bigfonts \bf }
 \vskip 20truept
\centertext{J.S.Dowker\footnote{ dowkeruk@yahoo.co.uk}} \vskip 7truept \centertext{\it
Theory Group,} \centertext{\it School of Physics and Astronomy,} \centertext{\it The
University of Manchester,} \centertext{\it Manchester, England} \vskip 7truept
\centertext{}

\vskip 7truept \vskip40truept
\begin{narrow}
GJMS operator determinants in odd dimensions are quickly computed for scalar and spinor
fields in both sub-- and super--critical cases as  a sum of Dirichlet eta functions with
polynomials in the (integer) operator order as coefficients.

\end{narrow}
\vskip 5truept
\vskip 60truept
\vfil
\end{title}
\pagenum=0
\newpage

\section{\bf1. Introduction}
Field theories on spheres for higher derivative propagation occur in an essential way in
connection with those AdS/CFT correspondences which have higher spins in the bulk.
Particular cases are the GJMS conformal scalars and their Dirac analogues. Formulas for the
effective action (`free energy') and conformal anomaly have been given  for free fields by
a purely spherical spectral method in [\pref{DowGJMS, DowGJMSspin}] and evaluated by
various means in [\pref{MandD}] and, later, by Brust and Hinterbichler, [\pref{BandH}].

If the order of the propagation operator exceeds a certain `critical' value, which depends
on the dimension, the operator ceases to exist, except in odd dimensions where a
continuation can be made. In this case the effective action acquires an imaginary part,
[\pref{DowGJMS}], essentially because of the existence of negative modes. Some values
are given in [\pref{BandH}] for particular dimensions.

More recently, Basile {\it et al}, [\pref{BJLL}], have given the explicit result (for
dimension equal to three) of a continuation of the free energy for any integral derivative
order, including above critical, in the cases of the higher order Rac and Di representations
(equivalent to GJMS scalars and spinors). In the present short, technical note, using a
quite different method, I extend the evaluation so as to apply easily to any given, odd
dimension.

In the next section I treat GJMS scalars (Rac$_k$) and then pass on to spinors (Di$_l$).
\section{\bf2. Scalar determinants}

I denote the scalar GJMS propagation operator of order $2k$ by $P^{Rac}_{2k}$. On the
$d$--sphere it is given by the Branson product,
  $$\eqalign{
  P^{Rac}_{2k}
  &=\prod_{j=1}^{k}\big(B_d^2-(j-1/2)^2\big)\cr
  &={\Ga\big(B_d+1/2+k\big)\over\Ga\big(B_d+1/2-k\big)}\,.
  }
  \eql{bgjms}
  $$

For scalars, $B_d= \sqrt{Y_d+1/4}$ where $Y_d$ is the conformally covariant
Penrose--Yamabe Laplacian. $k$ is, initially, an integer but the second expression allows
it to be extended to the reals. I will not consider this here.

To save time, I quote the expression derived in [\pref{DowGJMS}] for the scalar effective
action on the basis of the spectral data for odd spheres. It is

$$\eqalign{
  \log\det P^{Rac}_{2k}&={2(-1)^{d}\over d!}\int_0^k dz\,\,\pi z\,
  \tan\pi z\,\prod_{j=1}^{(d-1)/2}(z^2-(j-1/2)^2)\cr
  &\equiv {2(-1)^{d}\over d!}\int_0^k dz\,P(d,z)\,\pi\,\tan\pi z
  \,.
  }
  \eql{derivodd2}
  $$

The situation as $k$ increases is described in [\pref{DowGJMS}] section 8. I recapitulate a
little here. The polynomial $P$ cancels the poles in $\tan \pi z$ up to a certain value of
$z$. The first pole in the integrand appears at $z=d/2$ and so, if $k>d/2$, the integral is
undefined (infinite). However it can be rescued by extending $z$ into the complex plane
and running the integration along the real axis up to $k$, avoiding any poles. Since, it
turns out, the residues at the poles are integers, it doesn't matter how the poles are
skirted. The determinant remains unchanged.

The integral thus splits into a real part, given by its principal value, and an imaginary part
coming from the poles. I concentrate first on the real part as the harder to find.

I take $k$ to be integral. If it wasn't, then a further, numerical quadrature would be
required to make up the difference.

The integral can be broken up into unit pieces and a convenient change of variables then
yields,
  $$\eqalign{
  &\int_0^k dz\,P(d,z)\,\pi\,\tan\pi z
  =\sum_{i=0}^{k-1}\int_{-1/2}^{1/2}dx\,P(d,x+i+1/2)\,\pi\,\cot\pi x
  \,.
  }
  \eql{deriv}
  $$

Because of the oddness of the $\cot$, the finite part of the integral is obtained by
retaining just the odd powers of $x$ in $P$, say,
   $$
  P(d,x+i+1/2)=\sum_{p=0}^{(d-1)/2}K^p_d(i)\, x^{2p+1}+{\rm even\,powers} \,,
   $$
where the constants, $K_d^p(i)$ are polynomials in $i$ so allowing the sum over $k$ to
be done to give polynomials in $k$.

Hence the calculation is reduced to finding the integral, ($p\ge0$),
  $$\eqalign{
   C(p,x)&\equiv\pi\int_0 dx \,x^{2p+1}\,\cot(\pi x)\cr
   &=x^{2p+1}\,\log\sin(\pi x)-(2p+1)\int_0 dx \,x^{2p}\,\log\sin(\pi x)\,,
   }
   \eql{cotint}
  $$
evaluated at $x=1/2$, and I spend a little time on this evaluation. It can, of course, be
found in each case from a CAS but it is more satisfying to have a derivation from first
principles and a general formula.\footnote{ There is some interest in computing the
integral (\peq{cotint}) for any $x$, being related to polylogarithms (higher Spence
functions). An approach which depends on repeated integrations of $\cot\pi x$, or, more
conveniently, on $2p$ integrations of $\log\sin\pi x$, \ie of the Clausen integral will be
presented at another time.}

In exactly the case when $x=1/2$, an explicit expression is given by Crandall and Buhler,
[\pref{CandB}], obtained by expanding the power of $x$ in terms of Clausen functions,
and using the rather particular relation (a consequence of trigonometry)
  $$
  \ze(s)=2\int _0^{1/2}dx\cot\pi x\,S(s,x)\,,
  \eql{candb}
  $$
where  $S$ is the Clausen function
  $$
  S(s,x)=\sum_{n=1}^\infty {\sin(2\pi nx)\over n^s}\,,
  $$
which, for $s$ an odd integer, is a Bernoulli polynomial in $x$,
  $$
  S(s,x)=(-1)^{(s-1)/2} 2^{s-1}\pi^s{B_s(x)\over s!}\,.
  $$
Since the details in [\pref{CandB}] are somewhat sketchy, I give my own version here.

The first step is to introduce the twisted (or fermionic or alternating) Clausen function,
 $$
  \wt S(s,x)=\sum_{n=1}^\infty (-1)^{n-1}{\sin(2\pi nx)\over n^s}\,,
  $$
which is obtained, up to a sign, from $S(s,x)$ by making the central translation $x\to
x+1/2$,
  $$
  \wt S(s,x)=-S(s,x+1/2)\,.
  $$

This enables me without more ado to move directly to the Dirichlet eta function because
one now has,
  $$
  \eta(s)=2\int _0^{1/2}dx\cot\pi x\,\wt S(s,x)\,.
  \eql{candb2}
  $$

The polynomial form of the twisted Clausen function is, for $s$ odd,
  $$\eqalign{
  \wt S(s,x)&=-(-1)^{(s-1)/2} 2^{s-1}{\pi^s\over s!}B_s(x+1/2)\cr
  &\equiv(-1)^{(s+1)/2} {\pi^s\over 2\,s!}D_s(x)\,.\cr
  }
  \eql{polyf}
  $$

Next, the monomial $x^{\nu}$ in (\peq{cotint}) is expanded in the polynomials,
[\peq{candb}]. I do this in the following accelerated way.

The N\"orlund polynomials, $D_\nu(x)$, can also be expressed in terms of the central
derivative,
  $$
  D_\nu(x)={D\over2}\cosech{D\over2}\,\cdot (2x)^\nu\,.
  $$
which shows that $D_\nu(x)$ is an even polynomial if $\nu$ is even and an odd one if
$\nu$ is odd.
 The inversion,
  $$
  (2x)^\nu={\sinh D/2\over D/2}\,D_\nu(x)\,,
  $$
provides the required expansions using the basic relation
  $$
   D\,D_\nu(x)=2\nu\,D_{\nu-1}(x)\,.
  $$

The result is, for both odd and even $\nu$,
  $$
  x^\nu={2^{-\nu}\over \nu+1}\sum_{\mu=1,3,5\ldots}^{\nu,\nu+1}
  \comb{\nu+1} \mu\,D_{\nu-\mu+1}(x)\,,
  \eql{expn}
  $$
the upper limit being odd. For my purposes, I need only odd $\nu,=2p+1,$ and
(\peq{expn}), combined with (\peq{candb2}) and (\peq{polyf}) gives a quick derivation of
Crandall and Buhler's formula for (\peq{cotint}) at $x=1/2$,
  $$
  C(p,1/2)={(2p+1)!\over2^{2p+1}}\sum_{\mu=1,3,\ldots}^{2p+1}
  {(-1)^{{(\mu-1)}/2}\over(2p+2-\mu)!}{\eta(\mu)\over\pi^{\mu-1}}\,.
  \eql{ceep}
  $$

I now return to the expression, (\peq{deriv}) needed for calculating the (real part of) the
effective action,
  $$\eqalign{
  &\sum_{i=0}^{k-1}\sum_{p=0}^{(d-1)/2}\int_{-1/2}^{1/2}
  dx\,K^p_d(i) x^{2p+1}\pi\,\cot\pi x\cr
  &=\sum_{p=0}^{(d-1)/2}
L^p_d(k) \,C(p,1/2)\,,\cr
  }
  \eql{deriv2}
  $$
where,
  $$
  L_d^p(k)=2\sum_{i=0}^{k-1}K^p_d(i)\,,
  $$
is a polynomial in $k$. I will not spend time producing a formula for these polynomials in
Bernoullian terms and, since all quantities are now easily computable, I just list some
particular expressions of $\log\det P_{2k}$, for $d=3,5,7$ and 9 (remembering that
$\eta(1)=\log\,2$),

  $$
  {k(4k^2-1^2)\over12}\,\log\,2-{3k\over8}{\ze(3)\over\pi^2}\,,
  $$

$$
\frac{k\,\left( 4\,k^2-3^2\right) \,\left( 4\,k^2
-1^2\right)}{960}\,{\log}\,2 \,
 -{k(4k^2-3)\over16}{\zeta\left( 3\right)\over\pi^2}
+\frac{15\, \,k}{128}{\zeta\left( 5\right)\over\pi^4}\,,
$$

  $$\eqalign{
 {k (4k^2-5^2)(4k^2-3^2)(4k^2-1^2)\over161280}{\log}\,2
 &-{k(48k^4-200k^2+111)\over15360}{\ze(3)\over\pi^2}\cr
  &+{5k(4k^2-5)\over1024}{\zeta\left( 5\right)\over\pi^4}
   -\frac{63\,k\zeta\left( 7\right)}{2048\,{\pi }^{6}}\,,
   }
  $$

  $$\eqalign{
 &
\frac{k\,\left( 4\,k^2-7^2\right) \,\left( 4\,k^2-5^2\right) \,\left( 4\,k^2-3^2\right)
\,\left( 4\,k^2-1^2\right)}{46448640}{\log}\,2 \,\cr
 &-{k(6k^6-105k^4+455k^2-410)\over161280}{\zeta\left( 3\right)\over\pi^2}
 +{k(6k^4-50k^2+65)\over12288}{\zeta\left( 5\right)\over\pi^4}\cr
  &\hspace{***********}-{21k(4k^2-7)\over16384}{\zeta\left( 7\right)\over\pi^6}
 +\frac{255\,k\zeta\left( 9\right)}{32768\,{\pi }^{8}}\,.
 }
  $$
The general $\log\,2$ term is
  $$
  {1\over 2d!}\,k\!\prod_{j=1}^{(d-1)/2}\big(k^2-(j-1/2)^2\big)\,.
  $$

Although eta functions appear more immediately, I have converted them to Riemann zetas
as this is how such formulae are usually presented.

The $d=3$ formula was obtained by Basile {\it et al} [\pref{BJLL}].

The above expressions are valid strictly only for $k$ integral. Evaluation at specific
integers produces agreement with the results derived, in a longer way, in [\pref{MandD}]
and, later, in [\pref{BandH}]. Brunt and Hinterblicher, [\pref{BandH}], also give
super--critical values but they are not expressed as above.

\section{\bf 3. The imaginary part}

As stated, in the super--critical case (\ie $k>d/2)$ the integral acquires an imaginary part
coming from the active poles. In the approach here the exact value depends on the choice
of $z$--contour. Two basic contours are one ($\caC_1$)  just below (or above) the real
axis and one ($\caC_2$) that runs alternately above and below the poles. $\caC_1$ yields
the sum of the residues and $\caC_2$ an alternating sum.

Each residue equals the number of negative modes associated with each of the $k$ factors
in the product, (\peq{bgjms}). These can be labeled by the variable $j$, which runs from
$0$ to $k-1$. I denote the residues by $\rho_j$. Actually there are no negative modes, or
a pole, for the first factor, $j=0$, corresponding to the poles being $k-1$ in number. It is
convenient to distinguish `Dirichlet' and `Neumann' modes \ie those modes which are,
respectively, odd and even across the equator of the $d$--sphere, and write the number of
negative modes per factor as the $D$ and $N$ sum,
  $$\eqalign{
  \rho_j&=\rho^N_j+\rho^D_j\cr
  &=\rho^N_j+\rho^N_{j-1}\,,\cr
  }
  \eql{res2}
  $$
where the second equality is occasioned by the specific $D,N$ mode structures.

Either by mode counting in the way described in [\pref{DowGJMS}], leading to Ehrhart
polynomials, or from the explicit Plancherel  form of the integrand, it follows that
  $$
  \rho^N_j=\comb {(d-1)/2+j}d\,.
  \eql{res}
  $$

The contour $\caC_1$ then leads to the sum of the residues,
  $$\eqalign{
         N_T(k)\equiv&\sum_{j=1}^{k-1}\bigg[\comb {(d-1)/2+j}d+\comb {(d-3)/2+j}d\bigg]\cr
         &=\comb {(d-1)/2+k}{d+1}+\comb {(d+1)/2+k}{d+1}
         }
  $$
by the {\it hockey stick} identity.\footnote{ This is typical in that  summing over the
individual factors to give a GJMS quantity increases the dimension by one in  sort of
holographic way. See [\pref{dowct}] Appendix B.} This result can be written
  $$
    N_T(k)={2\over (d+1)!}\prod_{n=0}^{(d-1)/2}(k^2-n^2)\,.
  $$

The imaginary part of the effective action is $\pi N_T$ which agrees with [\pref{BJLL}],
(4.36).

How the contour circulates the poles can be thought of as corresponding to different
choices of logarithmic branch for the individual factors in the operator product. There is no
obligation to choose the same branch for each and the alternating contour, $\caC_2$, is
another distinguished option. An alternating sum of the residues, taking the relation
(\peq{res2}) into account, collapses by t\'elescopage to the first term,
  $$
  N_A(k)=\rho_{k-1}-\rho_{k-2}+\ldots\pm\rho_0=\rho^N_{k-1}\,.
  $$

This quantity is the number of negative eigenvalues of the GJMS operator
$P_{2k}^{Rac}$ as follows from the gamma function form of $P_{2k}^{Rac}$,
[\pref{DowGJMS}]. The number $N_T(k)$ does not give this measure because of sign
cancellations between factors in the eigenvalues (which are products).

The alternating sum of the Dirichlet negative eigenvalues equals the number of Dirichlet
negative eigenvalues of the product GJMS operator. The same holds  for the Neumann
case.
\section{4. \bf The Dirac field}

The GJMS--like Dirac operator takes the form,
$$\eqalign{
  P_{2k}^{Di}\equiv
  &B\prod_{h=1}^{l}\big(B^2-h^2\big)
  =\prod_{h=-l}^{l}\big(B+h\big)\,,\cr
  &={\Ga\big(B+1/2+k\big)\over\Ga\big(B+1/2-k\big)}\,.
  }
  \eql{dbgjm2}
  $$

For Dirac spinors $B= (\nsl^{\,2})^{1/2}=|\nsl|$ with an overall sign factor of
$\nsl/|\nsl|$ being understood, [\pref{BandO}]. The parameter $k$ is a half--integer
$k=l+1/2$, $l=0,1,2,\ldots$. The value $l=0$ gives the ordinary Dirac case.

It is shown in [\pref{DowGJMSspin}] that the logdet of this operator is,
$$\eqalign{
  \log\det P^{Di}_{2k}&={\caS\over d!}\int_0^k dz\,\,\pi z\,
  \cot\pi z\,\prod_{j=1}^{(d-1)/2}(z^2-j^2)\cr
  &\equiv {\caS\over d!}\int_0^k dz\,P(d,z)\,\pi\,\cot \pi z
  \,.
  }
  \eql{derivodd3}
  $$
The factor $\caS$ (a power of 2) relates mostly to spin degeneracy. I henceforth drop it.

As before, the integral can be split into unit sized pieces, plus, this time, an initial bit,
  $$
  \int_0^{1/2} dz\,P(d,z)\,\pi\,\cot \pi z+\sum_{j=0}^{l-1}
  \int_{-1/2}^{1/2} dz\,P(d,z+j+1)\,\pi\,\cot \pi z\,,
  $$
which again involves the integrals $C(p,1/2)$, (\peq{cotint}). Computation yields the
explicit expressions for $d=3,5,7$ and $9$, as shorter examples,

  $$
    {k(k^2-1^2)\over6}\,\log\,2-{3k\over16}{\ze(3)\over\pi^2}\,,
  $$

  $$
  {k(k^2-2^2)(k^2-1^2)\over120}\log\,2-{k(2k^2-3)\over64}{\zeta(3)\over\pi^2}
  +{15k\over256}{\ze(5)\over\pi^4}\,,
  $$

 $$\eqalign{
{(k^2-3^2)(k^2-2^2)(k^2-1^2)k\over5040}\,\log\,2&
-{k(3k^4-20k^2+21)\over1920}{\ze(3)\over\pi^2}+\cr
&+{5k(k^2-2)\over512}{\ze(5)\over\pi^4}-{7k\over640}{\ze(7)\over\pi^6}\,,
}
$$

$$\eqalign{
&{(k^2-4^2)(k^2-3^2)(k^2-2^2)(k^2-1^2)k\over362880}\log\,2\!
-\!{k(6k^6-105k^4+455k^2-410)\over161280}{\ze(3)\over\pi^2}\cr
&\hspace{**********}+{k(6k^4-50k^2+65)\over12288}{\ze(5)\over\pi^4}-
{21k(2k^2-5)\over16384}{\ze(7)\over\pi^6}+
{255k\over65536}\,{\ze(9)\over\pi^8}\,.
}
$$

The general $\log\,2$ term is
  $$
 {1\over d!}\,{\prod_{j=1}^{(d-1)/2}(k^2-j^2)}\,\log\,2\,.
  $$

The imaginary part follows in the same way as the scalar case. The Dirac mode structure
shows that $h$ plays the same role as $j$ did before, labelling the factors and the poles
($h=0,1,\ldots l$). Also, the inequality for a negative mode is identical to the Dirichlet
scalar one with $j\rightarrow h$. The residues coming from the relevant factors in
$P_{2k}^{Di}$ are then

  $$
  \rho_h=\comb{(d-1)/2+h}d\,,
  $$
and the total number of negative modes {\it from the factors} is,
  $$\eqalign{
        N_T^{Di}(l)&=\sum_{h=0}^l \rho_h=\comb{(d+1)/2 +l}{d+1}\cr
        &={l+(d+1)/2\over(d+1)!}\,l\prod_{j=1}^{(d-1)/2}(l^2-j^2)
        }
  $$
yielding the Di$_l$ imaginary part $\pi N_T^{Di}(l)$ as in [\pref{BJLL}].

I note that there are no negative modes for the first two brackets $h=0$ and $h=1$.

\section{\bf 5. From $d$ to $d+1$}

Equation (\peq{derivodd2})  for the scalar determinant is the algebraic consequence of
performing the GJMS sum of $k$ determinants of second order operators in $d$
dimensions. By inspection, the integrand is proportional to the Plancherel measure on the
$(d+1)$ dimensional hyperbolic space, H$^{d+1}$.

Another way of seeing this is to observe that identifying (\peq{derivodd2}) with an
alternative (equivalent) construction of the GJMS  $\log\det$ yields the identity,
[\pref{DowGJMS}],
$$
  \log{\det\big[\big(Y_{d+1}+1/4\big)^{1/2}-k\big]
  \over\det\big[\big(Y_{d+1}+1/4\big)^{1/2}+
  k\big]}=\int_0^k dz\,P(d,z)\,\pi\,\tan\pi z\,,
  \eql{bulk}
  $$
and, by differentiating with respect to $k$, it is seen that the residues of the poles of the
integrand at $d/2+j$ ($j=0,\ldots,k-1$) are the degeneracies of the eigenlevels,
$(d/2+j)^2$, of the $(d+1)$ operator, $Y_{d+1}+1/4$, on S$^{d+1}$. This is confirmed
by (\peq{res}) with (\peq{res2}).

S$^{d+1}$ is the Cartan dual of H$^{d+1}$ and the residue statement is in accordance
with general theorems on Plancherel measures, \eg\ [\pref{Camporesi}].

Equation (\peq{bulk}) can be regarded as a $(d+1)$ dimensional relation, the ratio
corresponding to the two boundary conditions in the double trace computation, in AdS/CFT
language, \eg\ [\pref{DandD}]. The derivation here is a purely boundary one.

All these statements can be transcribed into the Dirac case.
\section{\bf 6. Conclusion}

Another, rather particular, technique has been presented of rapidly computing the
determinants of GJMS operators on spheres for any specified odd dimension as a sum of
Dirichlet eta functions. Given the Plancherel form of the effective action, it is basically an
algebraic method using only mild properties of special functions. Neither Lerch
transcendents nor derivatives of the Hurwitz \zf\ appear. These occur, in some numbers, in
Basile {\it et al} [\pref{BJLL}], see also Bae [{\pref{bae}], who start from a different
integral representation of the free energy (effective action), one which is similar to those
in [\pref{dowinterp}] where original references can be found.\footnote{ The basic
ingredient is a Bessel transform the use of which dates back to the very earliest
discussions of \zfs\ on spheres and used, on and off, since. It yields a representation of
the \zf\ like the heat--kernel one but involving the {\it wave--kernel} (obtained from the
square root of the propagating operator). This is, more or less, the degeneracy generating
function, or, group theoretically, the character. In the hyperboiic case the transform is
applied slightly differently but ultimately leads to similar expressions for the determinants
after regularisation.}
\begin {ignore}
\section{\bf Appendix}

Although not required for the present calculation, the evaluation of the integral $C(p,x)$,
for any $x$, is of some interest.

I present here an approach which depends on repeated integrations of $\cot\pi x$, or,
more conveniently, on $2p$ integrations of $\log\sin\pi x$, \ie of the Clausen integral.

These integrations can be found from the series form of the Clausen function. To save
effort, use can be made of the results of Newman, [\pref{Newman}], which I briefly
review.

He independently defines a function
  $$
   \Upsilon(x)= -\int_0 dx\,\log|\sin x|
  $$
(equivalent to the Clausen integral). (I have changed his notation),

Many properties and values are derived but, more relevant for present purposes, is the
repeated integral, defined by recursion,
  $$
      \Up^n(x)=\int_0 dx\,\Up^{n-1}(x)
  $$
so that $\Upsilon\equiv\Upsilon^2$. This integral can be obtained by perpetual integration
of the textbook Fourier series of $ \log2\sin x$, (\eg\ Bromwich [\pref{Bromwich}]),

In order to use existing results directly, I remove the $\pi$s and put them back later.
Required is
  $$\eqalign{
  I(p,x)\equiv&\int_0 dx \,x^{2p}\,\log\sin(x)=-\int_0 dx \,x^{2p}\,\Up^1(x)
  =-\int_0 dx \,x^{2p}\,d\Up^2(x)\cr
  &=-x^{2p}\Up^2(x)+2p\int_0x^{2p-1}\Up^2(x)\cr
   &=-x^{2p}\Up^2(x)+2p\int_0x^{2p-1}d\Up^3(x)\cr
     &=-x^{2p}\Up^2(x)+2p\, x^{2p-1}\Up^3(x)-2p(2p-1)\int_0x^{2p-2}\Up^3(x)\cr
      &=-x^{2p}\Up^2(x)+2p\, x^{2p-1}\Up^3(x)-2p(2p-1)\int_0x^{2p-2}d\Up^4(x)\cr
 &=-x^{2p}\Up^2(x)+2p\, x^{2p-1}\Up^3(x)-2p(2p-1)x^{2p-2}\Up^4(x)\cr
 &+2p(2p-1)(2p-2)\int_0x^{2p-3}d\Up^5(x)\cr
 &=-x^{2p}\Up^2(x)+2p\, x^{2p-1}\Up^3(x)-2p(2p-1)x^{2p-2}\Up^4(x)\cr
 &-2p(2p-1)(2p-2)x^{2p-3}\Up^5(x)+2p(2p-1)(2p-2)(2p-3)\int_0x^{2p-4}d\Up^6(x)\cr
   }
  $$

For a given $p$, the process stops when the power of $x$ in the integrand vanishes
yielding a finite sum of the $\Up^n$ from $\Up^2$ to $\Up^{2p+2}$.

Sum is
  $$\eqalign{
  &=-x^{2p}\Up^2+\sum_{i=1}^{2p}(-1)^{i+1}2p(2p-1)\ldots(2p-i+1)x^{2p-i}\Up^{i+2}\cr
  &=-x^{2p}\Up^2-\sum_{i=1}^{2p}(-1)^i{(2p)!\over(2p-i)! }\,x^{2p-i}\Up^{i+2}(x)\cr
  &=-\sum_{i=0}^{2p}(-1)^i{(2p)!\over(2p-i)! }\,x^{2p-i}\Up^{i+2}(x)\cr
 }
  $$

Examples, $p=1$
  $$\eqalign{
  &=-x^{2}\Up^2(x)+2x\Up^3(x)-2\Up^4(x)\cr
 }
 \eql{comb1}
  $$
Evaluate at $\pi/2$
  $$\eqalign{
  &=-{\pi^2\over4}\Up^2(\pi/2)+\pi\Up^3(\pi/2)-2\Up^4(\pi/2)\cr
 }
  $$

As mentioned, the $\Up^n(x)$ can be derived by perpetual integration of the Fourier
series for $\log2\sin x$. The results are conveniently contained in equns. (35) and (36) of
[\pref{Newman}]. For general $x$, the expressions involve a finite series of the $\Up^n$
plus Clausen functions but for special values of $x$, \eg\ $\pi$ and $\pi/2$, they reduce to
a finite set of terms. These results are given in [\pref{Newman}]. For example,
  $$\eqalign{
   \Up^2(\pi/2)&={\pi\over2}\log2\cr
   \Up^3(\pi/2)&={\pi^2\over8}\log2+{7\over16}S_3\cr
   \Up^4(\pi/2)&={\pi^3\over48}\log2+{\pi\over8}\,S_3\cr
   \Up^5(\pi/2)&=-\frac{31\,S_5}{256}+\frac{{\pi }^{2}\,S_3 }{32}
   +\frac{{\pi }^{4}\,{\log}\left( 2\right) }{384},
}
  $$

The combination (\peq{comb1}) gives
  $$
  \eqalign{
  I(1,\pi/2)&=-{\pi^3\over24}\log2+{3\pi\over16}S_3
  }
  $$
which. as a check, agrees with machine evaluation.

All the expressions are readily coded and yield the results very speedily.

The cotangent integral is given (reinserting the $\pi$s) by,
  $$
  C(p,1/2)={2p+1\over\pi^{2p+1}}\,I(p,\pi/2)
  $$

  \newpage
Therefore need
  $$\eqalign{
  &\int_0^x dz \,z^{p-1}\,\log\sin(\pi z)= \int_0^x dz \,z^{p-1}\,\log{e^{i\pi z}-e^{-i\pi z}\over2i}\cr
  &=\int_0^x dz \,z^{p-1}\,\big(\log{e^{2i\pi z}-1\over2i}+\log e^{-i\pi z})\cr
  &=\int_0^x dz \,z^{p-1}\,\big(\log(e^{2i\pi z}-1)-\log 2i-i\pi z\big)\cr
  &={1\over p}\int_0^x dz^{p}\,\log(e^{2i\pi z}-1)-\int_0^x dz \,z^{p-1}(\log 2i+i\pi z)\cr
  &={1\over p}\int_0^x dz^{p}\,\log(e^{2i\pi z}-1)-{x^p\over p}\log 2i
  -{i\pi x^{p+1}\over p+1}\cr
  &={1\over p}\bigg(x^p\,\log(e^{2i\pi x}-1)
  -2\pi i\int_0^x dz\,{z^p\,e^{2i\pi z}\over e^{2i\pi z}-1}\bigg)-{x^p\over p}\log 2i
  -{i\pi x^{p+1}\over p+1}\cr
  &={1\over p}\bigg(x^p\,\log(e^{2i\pi x}-1)-
  {1\over (2i \pi)^p}\int_1^{e^{2i\pi x}} du\,{\log^pu\over u-1}\bigg)-{x^p\over p}\log 2i
  -{i\pi x^{p+1}\over p+1}\cr
  &={1\over p}\bigg(x^p\,\log(e^{2i\pi x}-1)-
  {1\over (2i \pi)^p}\int_0^{e^{2i\pi x}} du\,{\log^pu\over u-1}\bigg)-{x^p\over p}\log 2i
  -{i\pi x^{p+1}\over p+1}+C\cr
  &={(-1)^p (p-1)!\over(2i\pi)^p}Li_{p+1}(e^{2\pi ix})-{x^p\over p}\log 2i
  -{i\pi x^{p+1}\over p+1}+{1\over p}x^p i\pi+C\cr
  &={(-1)^p (p-1)!\over(2i\pi)^p}Li_{p+1}(e^{2\pi ix})-{x^p\over p}\log (-2i)
  -{i\pi x^{p+1}\over p+1}+C\cr
  }
  \eql{inte}
  $$
where $C$ is a constant (independent of $x$)
  $$
  C={1\over p (2i \pi)^p}\int_0^1 du\,{\log^pu\over u-1}
  =-(-1)^p\,{(p-1)!\over(2i\pi)^p} \,Li_{p+1}(1)
  $$

$C$ disappears  on construction of the integral
  $$
\int_{-x}^x =\int_0^x-\int_0^{-x}
  $$

Defining $u=e^{2i\pi z}$ so that $du=2i\pi e^{2i\pi z}\,dz$, $z={1\over2 i\pi}\,\log(u)$
and using the polylogarithm definition,

$$\eqalign{
{Li} _{p+1}(z)&={\frac {z (-1)^{p}}{p!}}\int_{0}^{1}{\frac {\log
^{p}(t)}{1-tz}}dt\cr
&={\frac {(-1)^{p}}{p!}}\int _{0}^{z}{\frac {\log
^{p}(u/z)}{1-u}}du\cr
&={\frac {(-1)^{p}}{p!}}\int_{0}^{z}{\frac {\log
^{p}(u)}{1-u}}du-{\frac{(-1)^{p}}{p!}}\int_{0}^{z}{\frac {\log
^{p}(z)}{1-u}}du\cr
&={\frac {(-1)^{p}}{p!}}\int_{0}^{z}{\frac {\log
^{p}(u)}{1-u}}du+{\frac{(-1)^{p}}{p!}}\log
^{p}(z)\log(1-z)\cr
&={\frac {(-1)^{p}}{p!}}\bigg(\int_{0}^{x}{\frac {\log
^{p}(u)}{1-u}}du+(2\pi ix)^p \log(1-e^{2i\pi x})\bigg)\cr
&={\frac {(-1)^{p}(2i\pi)^p}{ p!}}\bigg(-{1\over(2i\pi)^p}\int_{0}^{x}{\frac {\log
^{p}u}{u-1}}du+x^p \log(e^{2i\pi x}-1)-x^p i\pi\bigg)\cr
}
$$

Compare with Wolfram Alpha for $p=1$.
  $$\eqalign{
  &i({\pi\over2} z^2+{1\over2\pi}Li_2(e^{2i\pi z})-z\,\log(1-e^{2i\pi z})
  +z\log\sin\pi z\cr
  &={i\over2\pi}Li_2(e^{2i\pi z})+{i\pi z^2\over2}-z\log(-2i\sin\pi x\,e^{i\pi x})
  +z\log\sin\pi z\cr
  &={i\over2\pi}Li_2(e^{2i\pi z})+{i\pi z^2\over2}-z\log(-2i\,e^{i\pi z})\cr
  &={i\over2\pi}Li_2(e^{2i\pi z})+{i\pi z^2\over2}-z\log(-2i)-i\pi z^2\cr
  }
  $$

Above is
  $$\eqalign{
&{(-1)^p (p-1)!\over(2i\pi)^p}Li_{p+1}(e^{2\pi iz})-{z^p\over p}\log 2i
  -{i\pi z^{p+1}\over p+1}\cr\cr
&=-{1\over2i\pi}Li_2(e^{2\pi iz})-z\log(-2i)-{i\pi z^2\over2}\cr
}
  $$

Note
  $$
  {Li}_s(\pm i) = -2^{-s} \,\eta(s) \pm i \,\beta(s)
  $$

Wolfram Alpha gives, \eg,
  $$
  \int_0^{1/2} dz \,\log \sin \pi z=-{1\over 2}\log2
  $$
\ie real.

Check from (\peq{inte}) for $p=1$, $x$=1/2. Use $\log(i)=i\pi/2$
  $$ \eqalign{
  &{(-1)^p (p-1)!\over(2i\pi)^p}Li_{p+1}(e^{2\pi ix})-{x^p\over p}\log 2i
  -{i\pi x^{p+1}\over p+1}+{1\over p}x^p i\pi
  -(-1)^p\,{(p-1)!\over(2i\pi)^p} \,Li_{p+1}(1)\bigg|_{p=1}\cr
  &={-1\over2i\pi}Li_{2}(-1)-{1\over 2}\log 2i
  -{i\pi \over 4\cdot2}+{1\over 2} i\pi+{1\over2i\pi}Li_2(1)\cr
  &={1\over2i\pi}\eta(2)-{1\over 2}\log 2i
  -{i\pi \over 4\cdot2}+{1\over 2} i\pi+{1\over2i\pi}\ze(2)\cr
  &={1\over2i\pi}{\pi^2\over12}-{1\over 2}\log 2i
  -{i\pi \over 4\cdot2}+{1\over 2} i\pi+{1\over2i\pi}{\pi^2\over6}\cr
  &=-{i\pi\over24}-{1\over 2}\log 2i
  -{i\pi \over 8}+{1\over 2} i\pi-{i\pi\over12}\cr
  &={3\over12}i\pi-{1\over 2}\log 2i=-{1\over2}\log2
  }
  $$

Zagier has, in cut $z$ plane,
$$
Li_2(z) = -\int_0^z \log(1 - u) {du\over u}\,, z\in C/(1,\infty)=
-\int_0^1 \log(1 - zu) {du\over u}
$$
which gives sum definition by expansion of log.

Above has,
$$
\eqalign{
{Li} _{2}(z)&={z }\int_{0}^{1}{\frac {\log
(t)}{1-tz}}dt\cr
}
$$

\end{ignore}
\newpage
 \vglue 20truept

 \noin{\bf References.} \vskip5truept
\begin{putreferences}
     \ref{SandT}{Skvortsov,E.D.  and Tran,T. {\it AdS/CFT in Fractional Dimensions}, ArXiv:
   \break 1707.00758.}
     \ref{BandH2}{Brust,C. and Hinterbichler,K. {\it Partially Massless Higher--Spin Theory II:
   One-Loop Effective Actions},  ArXiv:1610.08522.}
   \ref{BandH}{Brust,C. and Hinterbichler,K. {\it Free $\square^k$ scalar conformal field theory},
   {\it JHEP} {\bf 02 }2017) 066, ArXiv: 1607.07439}
   \ref{dowinterp}{Dowker,J.S. {\it On a--F dimensional interpolation},ArXiv:1708.07094.}
   \ref{CandB}{Crandall, R.E.  and Buhler, J.P. {\it On the evaluation of Euler sums}, {\it
   Experimental Math.} {\bf 3}  (1994) 275.}
    \ref{BJLL}{Basile,T., Joung, E., Lal, S. and Li, W. {\it Character Integral Representation of
    Zeta Function in AdS$_{d+1}$: II}, ArXiv: 1805.10092.}
    \ref{bae}{Bae, J-B., Joung, E. and Lal, S. {\it One-loop test of Free SU(N) Adjoint
    Holography}, {\it JHEP} {\bf 04} (2016) 061,   ArXiv:1603.05387.}
    \ref{DowGJMSspin}{Dowker,J.{\it Spherical Dirac GJMS operator determinants},
  \jpamt{48}{2015}{025401}, ArXiv:1310.556.}
  \ref{Camporesi}{Camporesi,R.{\it Harmonic Analysis and Propagators on Homogeneous
  Spaces}, \prp{196}{1990}{1}.}
    \ref{dowct}{Dowker, J.S., {\it R\'enyi entropy and $C_T$ for higher derivative
    free scalars and spinors on even spheres}, ArXiv:1706.01369.}
   \ref{Dowcen}{Dowker,J.S., {\it Central differences, Euler numbers and symbolic methods},
 \break ArXiv:1305.0500.}
 \ref{moller}{M{\o}ller,N.M. \ma {343}{2009}{35}.}
 \ref{BandO}{Branson,T., and  Oersted,B. \jgp {56}{2006}{2261}.}
  \ref{BaandS}{B\"ar,C. and Schopka,S. {\it The Dirac determinant of spherical
     space forms},\break {\it Geom.Anal. and Nonlinear PDEs} (Springer, Berlin, 2003).}
 \ref{EMOT2}{Erdelyi, A., Magnus, W., Oberhettinger, F. and Tricomi, F.G. {
  \it Higher Transcendental Functions} Vol.2 (McGraw-Hill, N.Y. 1953).}
 \ref{Graham}{Graham,C.R. SIGMA {\bf 3} (2007) 121.}
  \ref{Morpurgo}{Morpurgo,C. \dmj{114}{2002}{477}.}
      \ref{DandP2}{Dowker,J.S. and Pettengill,D.F. \jpa{7}{1974}{1527}}
 \ref{Diaz}{Diaz,D.E. {\it JHEP} {\bf 0807} (2008) 103.}
    \ref{DandD}{Diaz,D.E. and Dorn,H. {\it JHEP} {\bf 0705} (2007) 46.}
    \ref{AaandD}{Aros,R. and Diaz,D.E. {\it Determinant and Weyl anomaly of
     Dirac operator: a holographic derivation}, ArXiv:1111.1463.}
  \ref{CandA}{Cappelli,A. and D'Appollonio, \pl{487B}{2000}{87}.}
  \ref{CandT2}{Copeland,E. and Toms,D.J. \cqg {3}{1986}{431}.}
   \ref{Allais}{Allais, A. {\it JHEP} {\bf 1011} (2010) 040.}
     \ref{Tseytlin}{Tseytlin,A.A. {\it On Partition function and Weyl anomaly of
     conformal higher spin fields} ArXiv:1309.0785.}
     \ref{KPS2}{Klebanov,I.R., Pufu,S.S. and Safdi,B.R. {\it JHEP} {\bf 1110} (2011) 038.}
    \ref{CaandWe}{Candelas,P. and Weinberg,S. \np{237}{1984}{397}.}
     \ref{ChandD}{Chang,P. and Dowker,J.S. \np{395}{1993}{407}.}
 \ref{Steffensen}{Steffensen,J.F. {\it Interpolation}, (Williams and Wilkins,
    Baltimore, 1927).}
     \ref{Barnesa}{Barnes,E.W. {\it Trans. Camb. Phil. Soc.} {\bf 19} (1903) 374.}
    \ref{DowGJMS}{Dowker,J.S. {\it Determinants and conformal anomalies of GJMS operators
    on spheres}, \jpa{44}{2011}{115402}, ArXiv:1010.0566.}
    \ref{Dowren}{Dowker,J.S. \jpamt {46}{2013}{2254}.}
 \ref{MandD}{Dowker, J.S. and Mansour,T. {\it Evaluation of spherical GJMS determinants},
 {\it J. Geom. Phys.} {\bf97} (2015) 51, ArXiv:1407.6122.}
 \ref{GandK}{Gubser,S.S and Klebanov,I.R. \np{656}{2003}{23}.}
     \ref{Dow30}{Dowker,J.S. \prD{28}{1983}{3013}.}
     \ref{Dowcmp}{Dowker,J.S. \cmp{162}{1994}{633}.}
     \ref{DowGJMSE}{Dowker,J.S. {\it Numerical evaluation of spherical GJMS operators
     for even dimensions} ArXiv:1310.0759.}

\end{putreferences}

\bye